\documentclass[aps, prd, twocolumn, reprint, a4paper, amsmath,
  nofootinbib, eqsecnum, showpacs, superscriptaddress]{revtex4-1}
\usepackage{graphicx}
\def\prsla{Proc.\ R.\ Soc.\ London\ A\ }
\def\cqg{Class.\ Quantum\ Grav.\ }
\begin{document}
\title
{Dynamic black holes through gravitational collapse: \break
   Analysis of multipole moment of the curvatures on the horizon}
%
\author{Motoyuki Saijo}
\email[E-mail: ]{saijo@rikkyo.ac.jp}
%
\affiliation
{Department of Physics, Rikkyo University, Toshima, Tokyo 171-8501,
  Japan}
\affiliation
{Research Center for Measurement in Advanced Science, 
  Rikkyo University, Toshima, Tokyo 171-8501, Japan}
%
\received{31 March 2011}
\accepted{2 June 2011}
%
\begin{abstract}
We have investigated several properties of rapidly rotating dynamic 
black holes generated by gravitational collapse of rotating
relativistic stars. At present, numerical simulations of the binary
black hole merger are able to produce a Kerr black hole of $J_{\rm
  final} / M_{\rm final}^2$ up to $= 0.91$, of gravitational collapse
from uniformly rotating stars up to $J_{\rm final} / M_{\rm final}^2 
\approx 0.75$, where $J_{\rm final}$ is the total angular momentum and
$M_{\rm final}$ the total gravitational mass of the hole.  We have
succeeded in producing a dynamic black hole of spin $J_{\rm final} /
M_{\rm final}^2\approx$ $0.95$ through the collapse of differentially 
rotating relativistic stars.  We have investigated those dynamic
properties through diagnosing multipole moment of the horizon, and
found the following two features.  Firstly, two different definitions
of the angular momentum of the hole, the approximated Killing vector 
approach and dipole moment of the current multipole approach, make no
significant difference to our computational results.  Secondly,
dynamic hole approaches a Kerr by gravitational radiation within the
order of a rotational period of an equilibrium star, although the
dynamic hole at the very forming stage deviates quite far from a Kerr.
We have also discussed a new phase of quasi-periodic waves in the
gravitational waveform after the ringdown in terms of multipole moment
of the dynamic hole.
\end{abstract}
%
\maketitle
%
\section{Introduction
\label{sec:intro}}
There are various mass ranges of black holes (BHs) in nature.
Supermassive BHs exist in the center of most galaxies, and the typical
mass range of this category is around $10^{5} M_{\odot}$ -- $10^{10}
M_{\odot}$.  In spite of clear evidence of its existence, the actual
formation scenario for the supermassive BHs is still not certain
\citep{Rees03}.  There are several candidates for the intermediate
mass range (around $10^{2} M_{\odot}$ -- $10^{3} M_{\odot}$) of the
BHs in the globular clusters \cite{MC04}.  At present the object of
the intermediate mass range has not been directly found yet.
Moreover, the standard formation scenario for such an object has to
pass through the formation of stellar mass objects.  The merger of
stellar holes or compact objects, or collision and collapse of massive
stars is a typical scenario for forming such intermediate mass range
objects.  There are also some other candidates for stellar mass range
(around $3 M_{\odot}$ -- $50 M_{\odot}$) of BHs in our galaxy.  Binary
coalescence of the stars or the collapse of the star of stellar mass
range is a typical scenario for forming such objects.

Nowadays, we can produce a dynamic BH by computer.  We have two
representative scenarios for forming a dynamic BH promptly.  Here we
neglect accretion, since the standard timescale for this process is
considerably longer than the dynamical one.  The first one is the
merger of equal mass, binary BHs.  Based on the current numerical
simulations of the binary BHs, there may exist an upper spin limit for
the newly formed BH.  The binary composed of non-spinning individual
BHs leads to the final maximum spin of the newly formed BH of $J_{\rm
  final}/M_{\rm final}^{2} = 0.69$ ($J_{\rm final}$ is the angular
momentum and $M_{\rm final}$ the gravitational mass of the newly
formed BH) \citep{SBCKMP09}, while the spinning individual BHs in
arbitrary direction leads to a final BH spin of $J_{\rm final}/M_{\rm
  final}^{2} = 0.91$ \citep{MTBGS08}.  Moreover, test particle
approximation in BH perturbation approach including the superradiance 
effect leads to a final BH spin of $J_{\rm final}/M_{\rm final}^{2} =
0.9979$ \citep{KLP10}.  In theory, we have the following discussion to
support the existence of the upper spin limit of the newly formed BH.
If the plunge phase of the binary BHs is characterised by the physical
quantities at a certain separation radius, namely the ISCO (innermost
stable circular orbit) of a newly formed BH, then there may exist an
upper limit to the spin of the newly formed BH because most likely
there exist a radially unstable condition at the ISCO under which the
binary begins to collide and form a new BH.

The next one is gravitational collapse of a uniformly rotating
relativistic star.  In this scenario, the maximum spin of the BH
exists by the following discussion.  First, a star contracts itself to
the mass shedding limit, conserving the angular momentum of the
system.  Then, so far as the system contains sufficient angular
momentum, the star evolves along the mass shedding sequence
quasi-stationary, releasing the mass and angular momentum.  Once the
star reaches the critical onset of collapse because of relativistic
gravitation, it begins to collapse \citep{ZN96}.  From the collapse of
a uniformly rotating supermassive star, the final spin of a newly
formed BH is around $J_{\rm final}/M_{\rm final}^{2} \approx 0.75$
\citep{SS02}.

\begin{table*}[htbp]
\begin{center}
\caption{
Two different radially unstable rotating equilibrium supermassive
stars for a BH formation} 
\begin{ruledtabular}
\begin{tabular}{c c c c c c c}
Model & 
$R_{\rm p} / R_{\rm e}$\footnotemark[1] & 
$\rho_{0}^{\rm max}$\footnotemark[2] & 
$M$\footnotemark[3] &
$T/W$\footnotemark[4] &
$J/M^2$\footnotemark[5] & 
$M/R$\footnotemark[6]
\\
\hline
  I & $0.450$ & $1.56 \times 10^{-5}$ & $4.88$ &
  $0.108$ & $0.99$ & $2.56 \times 10^{-2}$
\\
 II & $0.425$ & $1.56 \times 10^{-5}$ & $5.07$ & $0.118$ & $1.03$
 & $2.63 \times 10^{-2}$
\\
\end{tabular}
\end{ruledtabular}
\label{tab:equilibrium}
\footnotetext[1]{Ratio of the polar proper radius to the equatorial
  proper radius}
\footnotetext[2]{Maximum rest mass density}
\footnotetext[3]{Gravitational mass}
\footnotetext[4]{Ratio of the rotational kinetic energy to the
  gravitational binding energy}
\footnotetext[5]{$J$: Total angular momentum}
\footnotetext[6]{$R$: Circumferential radius}
\end{center}
\end{table*}

\begin{figure}
\centering
\includegraphics[keepaspectratio=true,width=8cm]{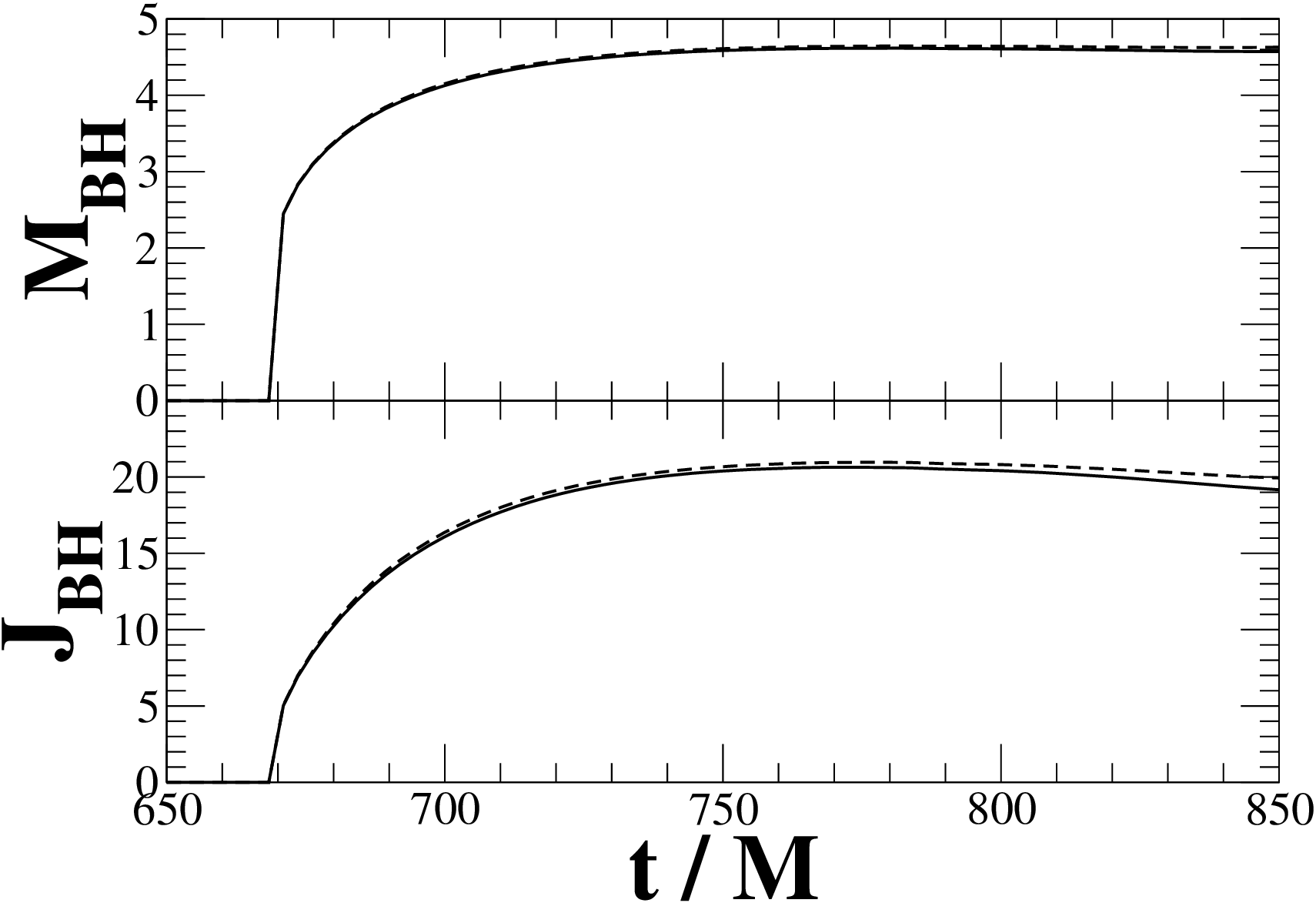}
\caption{
The gravitational mass and angular momentum of the dynamic BH through
gravitational collapse for model I.  The angular momentum is computed
by the approximate Killing vector (solid line), while by dipole moment
of the imaginary part of the Weyl scalar $\Psi_{2}$ (dashed line) on
the apparent horizon.  The gravitational mass is computed by using the
first law of BH thermodynamics for both cases.
\label{fig:mj1}
}
\end{figure}

\begin{figure}
\centering
\includegraphics[keepaspectratio=true,width=8cm]{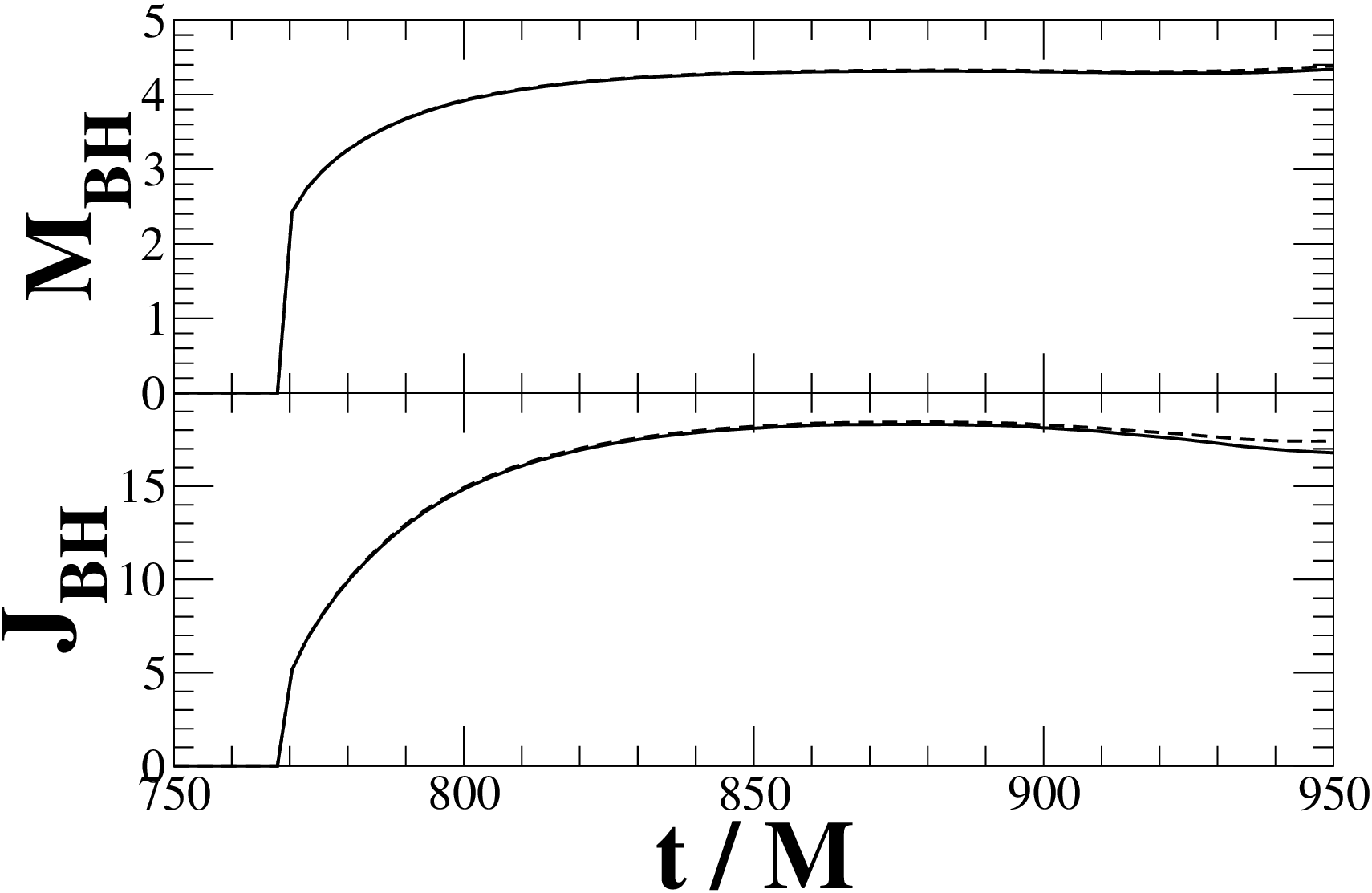}
\caption{
Same as Fig.~\ref{fig:mj1} but for model II.
\label{fig:mj2}
}
\end{figure}

\begin{figure}
\centering
\includegraphics[keepaspectratio=true,width=8cm]{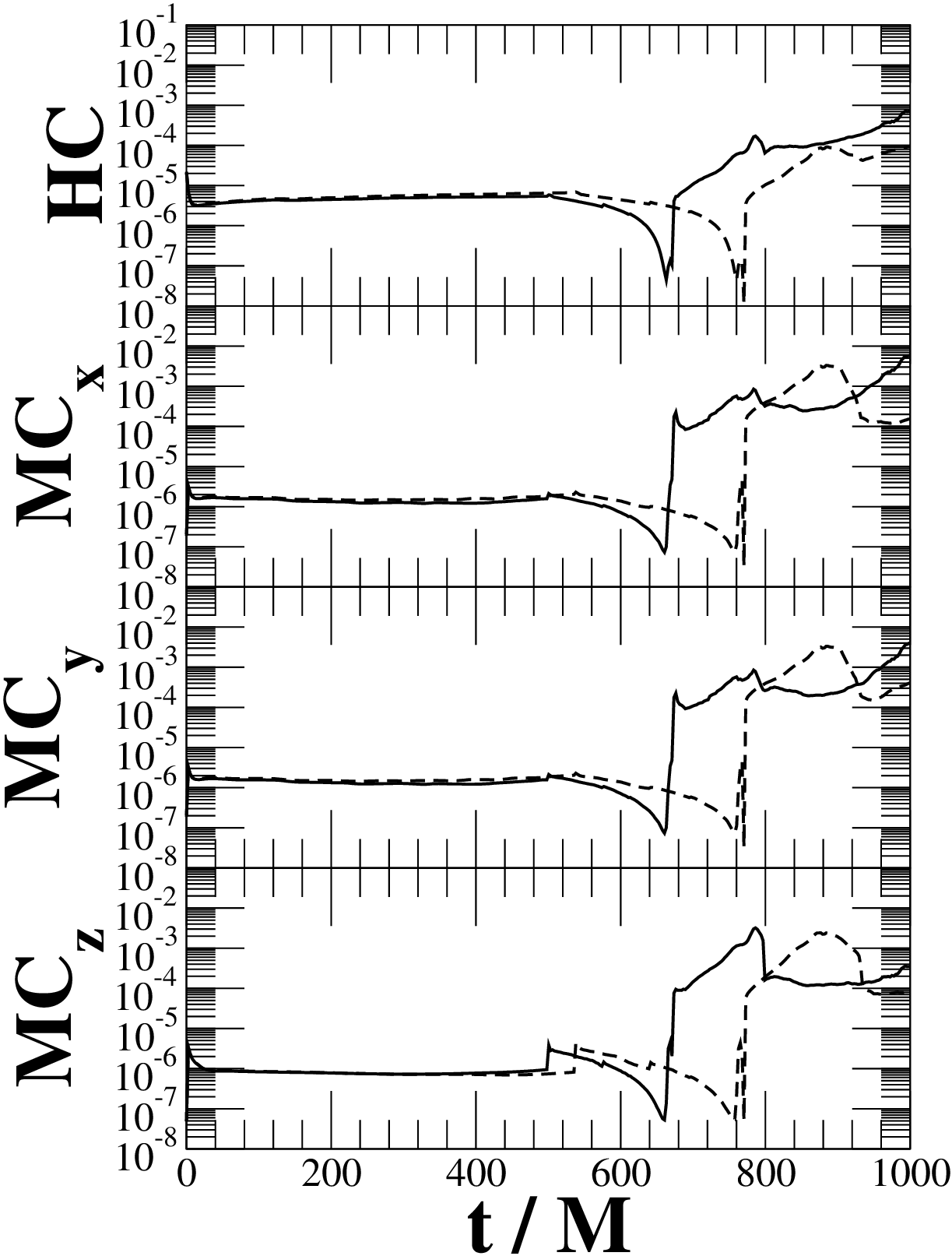}
\caption{
Euclidean norm of the normalised Hamiltonian constraint and the
normalised Momentum constraints throughout the evolution for models I
(solid line) and II (dashed line).
\label{fig:const}
}
\end{figure}

One of the primary observational missions for detecting gravitational
waves  in ground-based and space-based interferometers is to
investigate a various mass range of BHs and compact objects
\citep{Thorne98}.  Combining the global network of gravitational wave
detectors, we are in these decades able to extract fruitful features
of BHs in the frequency band of $10^{-4} {\rm Hz}$ -- $10^{3} {\rm
  Hz}$.  Potential sources of high signal to noise events in this
frequency range are quasi-periodic waves arising from nonaxisymmetric
bars in collapsing relativistic stars and from the inspiral of binary
BHs for example (e.g.\ Ref.~\citep{SS09}).  In addition, a
nonspherical collapse of a rotating relativistic star to a BH
potentially generates a significant amount of burst waves and
quasi-normal ringing waves (e.g.\ Ref.~\citep{FN03}).  In this paper
we trace the collapse of relativistic stars through numerical
simulations to investigate some of these possibilities. 

Here we relax the condition of uniformly rotating profile in the
equilibrium to produce a highly spinning dynamic BH.  Differential
rotation profile of the star enables us to impose large amount of
angular momentum in the system, since it relaxes the restriction to
the angular velocity at the equatorial radius, which comes from the
limitation of the mass-shedding.  According to the above idea,
\citet{SH09} have succeeded in producing a dynamic BH of spin $J_{\rm
  final}/M_{\rm final}^{2} \approx 0.98$.  Here we focus on the BH
configuration in this paper by using multipole moment of the
curvatures on the apparent horizon.  We try to answer the following
questions.  Can we extract precisely the mass and angular momentum of
a dynamical BH by using multipole moment of the curvatures on the
horizon?  Can a newly formed BH be represented as a stationary Kerr BH
at several dynamical times after the BH formation?  Is it useful to 
use multipole moment of a dynamic BH to extract some properties of a
BH, and to find a cause of quasi-periodic gravitational waves after
the ringdown, for example?  To answer these questions, three spatial 
dimensional general relativistic hydrodynamics is necessary.

The content of this paper is as follows.  In Sec.~\ref{sec:bequation},
we briefly explain the general relativistic hydrodynamics, especially
the numerical tools we use to understand the property of a dynamic
BH.  In Sec.~\ref{sec:NResults}, we introduce our findings of a
dynamic BH, focusing on its configuration.
Section~\ref{sec:Conclusions} is devoted to the summary of this
paper.  Throughout this paper, we use the geometrized units with
$G=c=1$ and adopt Cartesian coordinates $(x,y,z)$ with the coordinate
time $t$.  Note that Greek index takes $(t,x,y,z)$, while Latin one
takes $(x,y,z)$.

\section{Basic tools in numerical relativity
\label{sec:bequation}}
In this section, we briefly describe three-dimensional relativistic
hydrodynamics in full general relativity.  We also explain our
techniques for investigating outgoing gravitational waves from the
sources and a dynamic horizon configuration (see Ref.~\citep{SH09} and
references cited therein).

\subsection{The gravitational field equations}

We define a spatial projection tensor $h^{\mu\nu} \equiv g^{\mu\nu}
+ n^{\mu} n^{\nu}$, where $g^{\mu\nu}$ is the spacetime metric,
$n^{\mu} = (1/\alpha, -\beta^i/\alpha)$ the unit normal to the spatial
hypersurface, and where $\alpha$ and $\beta^i$ are the lapse and
shift.

We evolve the spacetime with the 17 spacetime associated variables
($\phi$, $K$, $\tilde{\gamma}_{ij}$, $\tilde{A}_{ij}$,
$\tilde{\Gamma}^{i}$), where $e^{\phi}$ is the conformal factor,
$K_{ij}$ the extrinsic curvature, $\tilde{\gamma}_{ij}$ the
conformally related spatial 3-metric, $\tilde{A}_{ij}$ the conformally
related trace-free extrinsic curvature, and $\tilde{\Gamma}^{i}$ the
conformal connection function.  The evolution equations are
\begin{widetext}
\begin{eqnarray}
\left( \frac{\partial}{\partial t} - {\cal L}_{\beta} \right) 
\phi &=& 
-\frac{1}{6} \alpha K, 
\label{eqn:evolution_phi}
\\
\left( \frac{\partial}{\partial t} - {\cal L}_{\beta} \right) 
K &=& 
- \gamma^{ij} D_{i} D_{j} \alpha 
+ \alpha \left[ \tilde{A}_{ij} \tilde{A}^{ij} +\frac{1}{3} K^{2} +
  \frac{1}{2} (\rho_{\rm H} + S) \right],
\\
\left( \frac{\partial}{\partial t} - {\cal L}_{\beta} \right) 
\tilde{\gamma}_{ij} &=&
- 2 \alpha \tilde{A}_{ij},
\\
\left( \frac{\partial}{\partial t} - {\cal L}_{\beta} \right) 
\tilde{A}_{ij} &=&
e^{-4\phi} [ -D_{i} D_{j} \alpha + \alpha (R_{ij} - S_{ij}) ]^{\rm TF}
+ \alpha ( K \tilde{A}_{ij} - 2 \tilde{A}_{il} \tilde{A}^{l}_{j} ),
\\
\left( \frac{\partial}{\partial t} - {\cal L}_{\beta} \right) 
\tilde{\Gamma}^i &=&
- 2 \tilde{A}^{ij} \frac{\partial}{\partial x^{j}} \alpha + 2 \alpha 
\left( 
  \tilde{\Gamma}^{i}_{jk} \tilde{A}^{jk} 
  - \frac{2}{3} \tilde{\gamma}^{ij} \frac{\partial}{\partial x^{j}} K 
  - \tilde{\gamma}^{ij} S_{j} 
  + 6 \tilde{A}^{ij} \frac{\partial}{\partial x^j} \phi
\right)
\nonumber \\
&&
- \frac{\partial}{\partial x^j}
\left(
  \beta^{l} \frac{\partial}{\partial x^l} \tilde{\gamma}^{ij} 
  - 2 \tilde{\gamma}^{m (j} \frac{\partial}{\partial x^{m}} \beta^{i)} 
  + \frac{2}{3} \tilde{\gamma}^{ij} \frac{\partial}{\partial x^{l}}
  \beta^{l}
\right),
\end{eqnarray}
\end{widetext}
where ${\cal L}_{\beta}$ denotes the Lie derivative along the shift
$\beta^{i}$, $\rho_{\rm H} = T_{\mu \nu} n^{\mu} n^{\nu}$, $S_i =
T_{\mu \nu} n^{\mu} h^{\nu}_{i}$ and TF the trace-free part of the
tensor.  This set of equations for solving the Einstein's field
equations numerically is usually called the BSSN formalism.  As for
gauge conditions, we choose the generalised hyperbolic $K$-driver
\citep{ABDKPST03} for the lapse, and the generalised hyperbolic
$\tilde{\Gamma}$-driver \citep{BCCKM06} for the shift.

\subsection{The matter equations}
We assume a perfect fluid for describing a relativistic star as 
\begin{equation}
T^{\mu \nu} = 
\rho \left( 1 + \varepsilon + \frac{P}{\rho} \right) u^{\mu} u^{\nu} +
Pg^{\mu\nu},
\end{equation}
where $\rho$ is the rest-mass density, $\varepsilon$ the specific
internal energy, $P$ the pressure, and $u^{\mu}$ the four-velocity.

Energy momentum conservation $\nabla_{\mu} T^{\mu\nu}=0$ together with
a continuity equation, leads to the flux conservative form of the
relativistic continuity, the relativistic energy and the relativistic
Euler equations as \citep{BFIMM97}
\begin{equation}
\frac{1}{\sqrt{-g}} \frac{\partial}{\partial t} 
  (\sqrt{\gamma} {\boldsymbol{\cal U}}) + 
\frac{1}{\sqrt{-g}} \frac{\partial}{\partial x^i} 
  (\sqrt{-g} {\boldsymbol{\cal F}}^i)
= \boldsymbol{\cal S}^i,
\end{equation}
where the state vector $\boldsymbol{\cal U}$, the flux vectors
$\boldsymbol{\cal F}^i$, and the source vectors $\boldsymbol{\cal
  S}^i$ are
\begin{widetext}
\begin{eqnarray}
\boldsymbol{\cal U} &=& 
  [D, S_i, \tau]^{T},\\
\boldsymbol{\cal F}^i &=&
  \left[ D \left( v^i - \frac{\beta^i}{\alpha} \right), 
   S_j \left( v^i - \frac{\beta^i}{\alpha} \right) + P \delta^{i}_{j},
   \tau \left( v^i - \frac{\beta^i}{\alpha} \right) + P v^{i}
  \right]^{T},\\
\boldsymbol{\cal S}^i &=&
  \left[0, 
    T^{\mu \nu} 
    \left( 
      \frac{\partial}{\partial x^{\mu}} g_{\nu j} -
      \Gamma^{\delta}_{\nu \mu} g_{\delta j}
    \right),
    \alpha 
    \left( 
      T^{\mu 0} \frac{\partial}{\partial x^{\mu}} \ln \alpha -
      T^{\mu \nu} \Gamma^{0}_{\nu \mu}
    \right)
  \right]^{T},
\end{eqnarray}
\end{widetext}
and $\Gamma^{i}_{~jk}$ is a Christoffel symbols.  Note that $(\rho,
v_{i}, \varepsilon)$ are the physical variables of the above
equations, and the flux conserved quantities $D$, $S_i$, $\tau$ are
\begin{eqnarray}
D &=& \rho W, \\
W &=& \alpha u^{t}, \\
S_{i} &=& \rho h W^{2} v_{i}, \\ 
E &=& \rho h W^{2} - P,\\
\tau &=& E - D,
\end{eqnarray}
where $v_{i} = u_{i} / W$, $h \equiv 1 + \varepsilon + P / \rho$ is
the specific enthalpy.  In the Newtonian limit, the above three
physical variables coincide with the rest mass density, the flux
density of the rest mass, and the energy of a unit volume of the
fluid.  In order to solve the set of equations, we have to impose an
additional condition among the thermodynamical quantities, namely the
equation of state.  We adopt a $\Gamma$-law equation of state in the
form 
\begin{equation}
P = (\Gamma - 1) \rho \varepsilon,
\label{gammalaw1}
\end{equation}
where $\Gamma$ is the adiabatic index which we set to $4/3$ in this
paper, representing a supermassive star (the pressure is dominated by
radiation). 

\begin{figure}
\centering
\includegraphics[keepaspectratio=true,width=8cm]{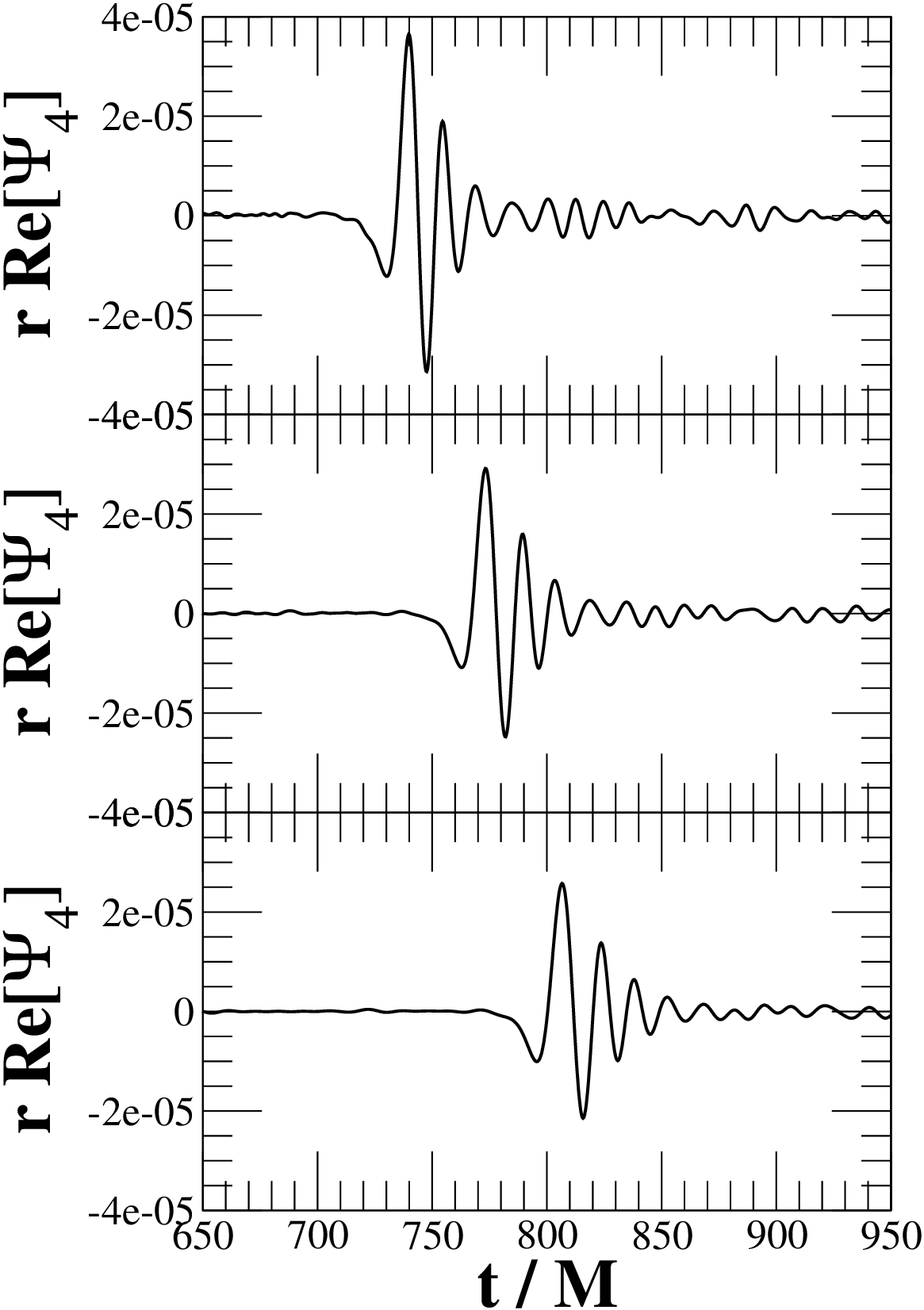}
\caption{
Gravitational waveforms for model I.  We monitor the real part of the
Weyl scalar $\Psi_{4}$, which represents the plus mode of outgoing
waves at null infinity.  The observer is located in the $x$-direction 
of the equatorial plane at $x=65.52 M$, $98.29 M$, $131.05 M$ from the 
top panel, respectively.  Note that the apparent horizon appears in
the hypersurface after $t = 670.97 M$.
\label{fig:gw1}
}
\end{figure}

\begin{figure}
\centering
\includegraphics[keepaspectratio=true,width=8cm]{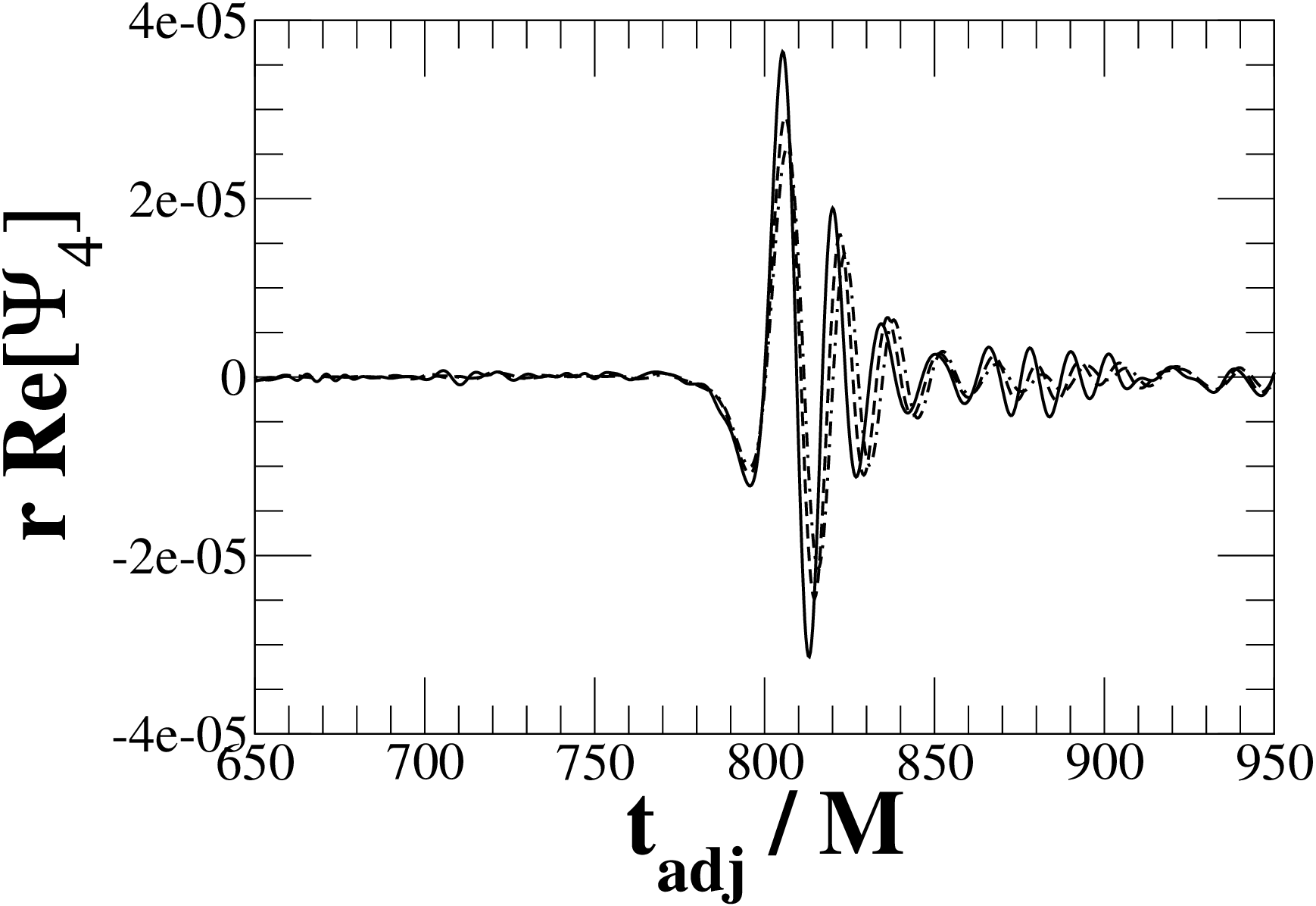}
\caption{
Same as Fig.~\ref{fig:gw1} but adjustment of time shift, plotted in
the same panel.  Solid, dashed and dash-dotted line denotes the
waveform detected in the $x$-direction of the equatorial plane at
$x=65.52 M$, $98.29 M$, $131.05 M$.  We use the following adjustment
of time shift $t_{\rm adj} \equiv t + x_{\rm farthest~observer} -
x_{\rm obs}$, assuming that gravitational waves propagate with the
speed of light in flat spacetime.
\label{fig:gw1comp}
}
\end{figure}

\begin{figure}
\centering
\includegraphics[keepaspectratio=true,width=8cm]{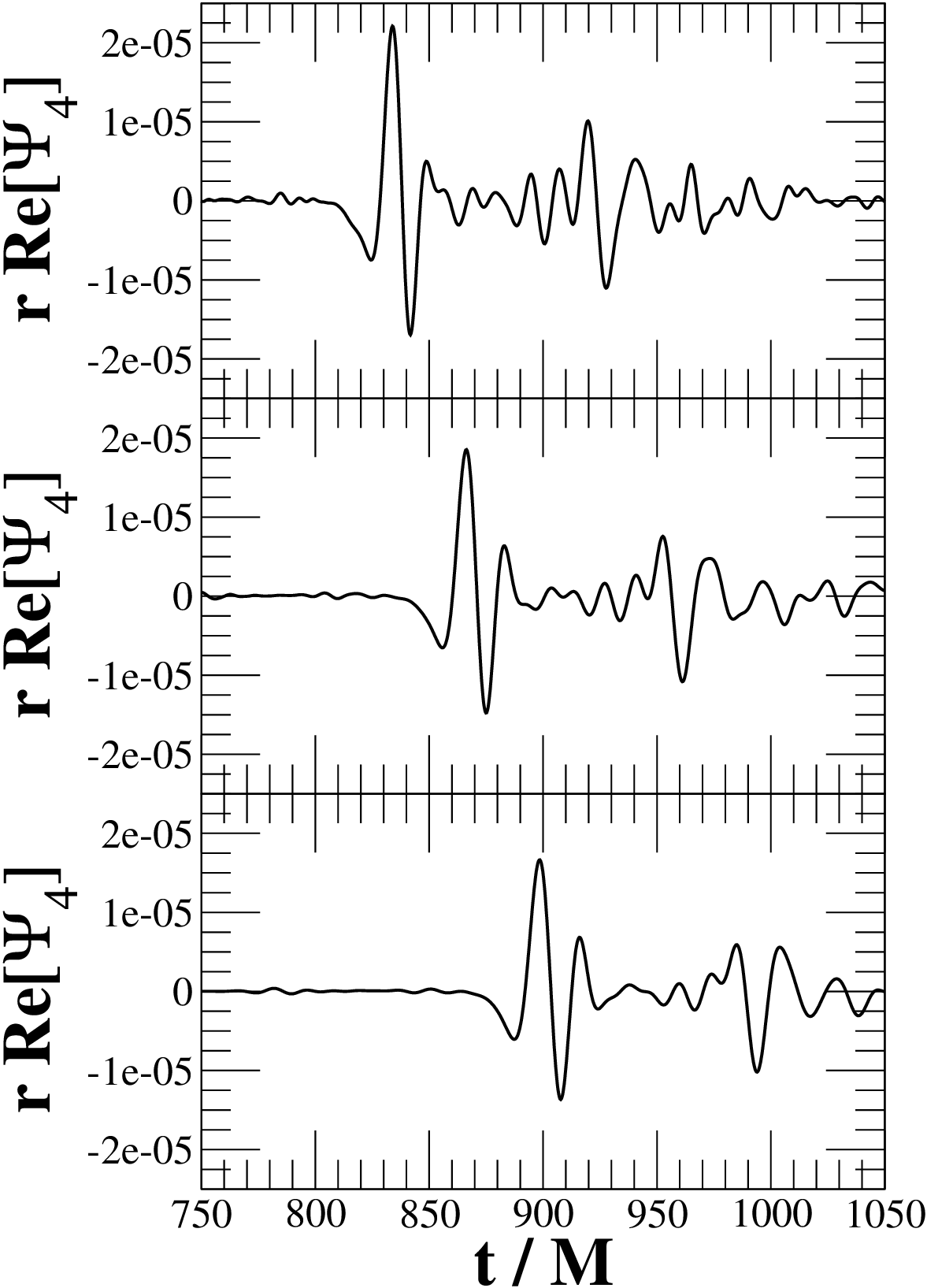}
\caption{
Same as Fig.~\ref{fig:gw1} but for model II.  The observer is located
in the $x$-direction of the equatorial plane at $x=63.15 M$, $94.72
M$, $126.30 M$ from the top panel, respectively.  Note that the
apparent horizon appears after $t = 770.40 M$.
\label{fig:gw2}
}
\end{figure}

\begin{figure}
\centering
\includegraphics[keepaspectratio=true,width=8cm]{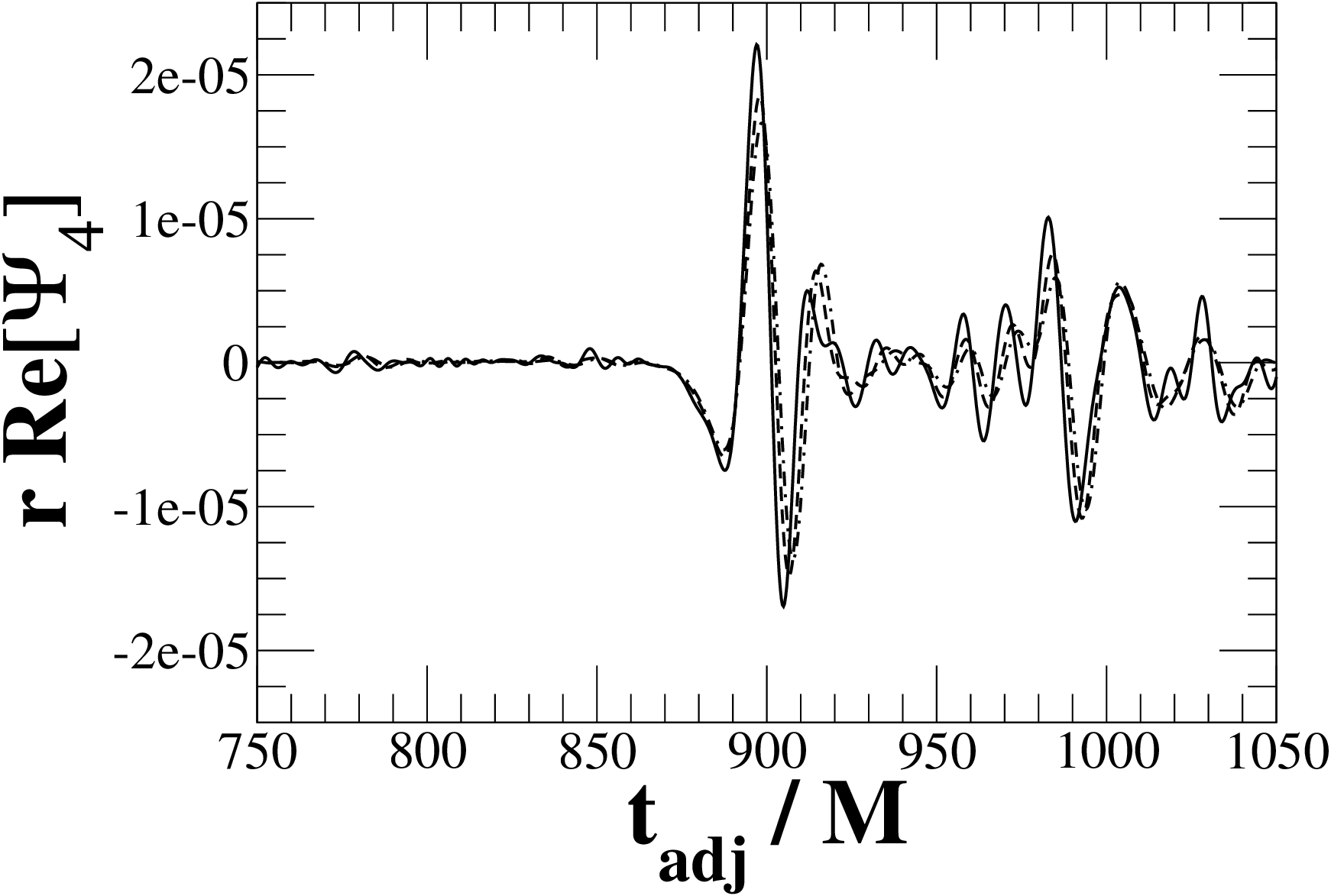}
\caption{
Same as Fig.~\ref{fig:gw2} but adjustment of time shift, plotted in
the same panel.  Solid, dashed and dash-dotted line denotes the
waveform detected in the $x$-direction of the equatorial plane at
$x=63.15 M$, $94.72 M$, $126.30 M$.  We use the following adjustment
of time shift $t_{\rm adj} \equiv t + x_{\rm farthest~observer} -
x_{\rm obs}$, assuming that gravitational waves propagate with the 
speed of light in flat spacetime.
\label{fig:gw2comp}
}
\end{figure}

\subsection{Gravitational waveforms}
We monitor the Weyl scalar $\Psi_4$ in Newman-Penrose formalism for
investigating outgoing gravitational waves as
\begin{equation}
\Psi_{4} = 
  C_{\mu \nu \lambda \sigma} k^{\mu} \bar{m}^{\nu} k^{\lambda} \bar{m}^{\sigma},
\end{equation}
where $C_{\mu \nu \lambda \sigma}$ is the Weyl tensor, $k^{\mu}$ the
ingoing null vector, $m^{\mu}$ and $\bar{m}^{\mu}$ are the orthogonal
spatial-null vectors of the four complex null tetrad ($l^{\mu}$,
$k^{\mu}$, $m^{\mu}$, $\bar{m}^{\mu}$).  The Weyl scalar $\Psi_4$
represents the outgoing gravitational waves at infinity
\begin{equation}
\Psi_{4} = \ddot{h}_{+} - i \ddot{h}_{\times},
\end{equation}
where $h_{+}$ and $h_{\times}$ are the two polarisation modes
(transverse-traceless condition) of the perturbed metric from flat
spacetime in spherical coordinate, and $\dot{q}$ is the time
derivative of the quantity $q$.  The Weyl scalar $\Psi_4$ roughly
represents the outgoing gravitational waves, ignoring the radiation
scattered back by the curvature when locating the observer far from
the source.  Therefore we trace the Weyl scalar $\Psi_4$ to understand
key features of gravitational waves emitted from this system.

\begin{figure}
\centering
\includegraphics[keepaspectratio=true,width=8cm]{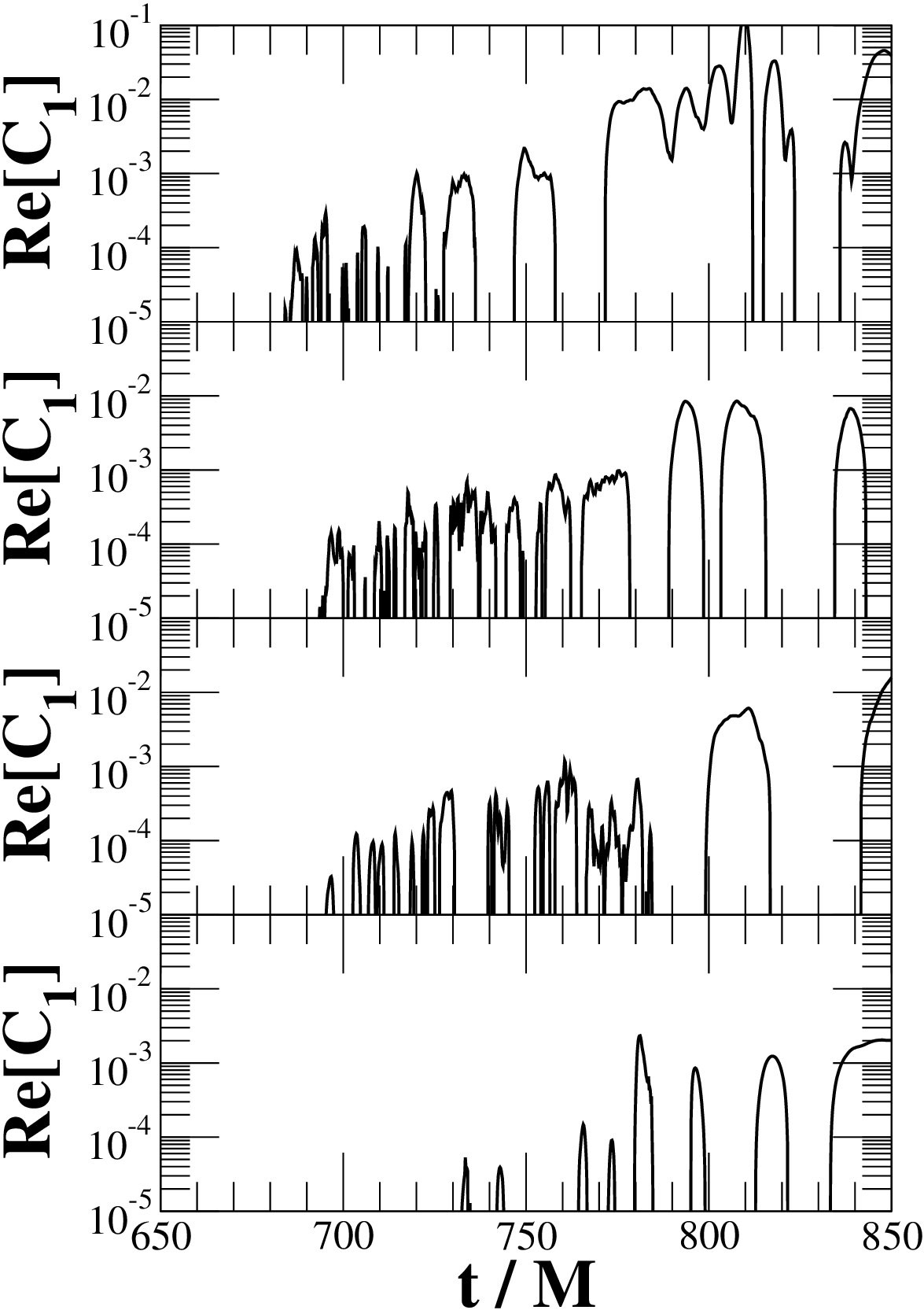}
\caption{
The $m=1$ diagnostics of the rest mass density along the equatorial
ring for model I.  We measure the diagnostics in the equatorial plane
at the radius $r=1.024 M$, $2.048 M$, $4.095 M$, $10.24 M$ from the
top panel, respectively.
\label{fig:md1rm1}
}
\end{figure}

\begin{figure}
\centering
\includegraphics[keepaspectratio=true,width=8cm]{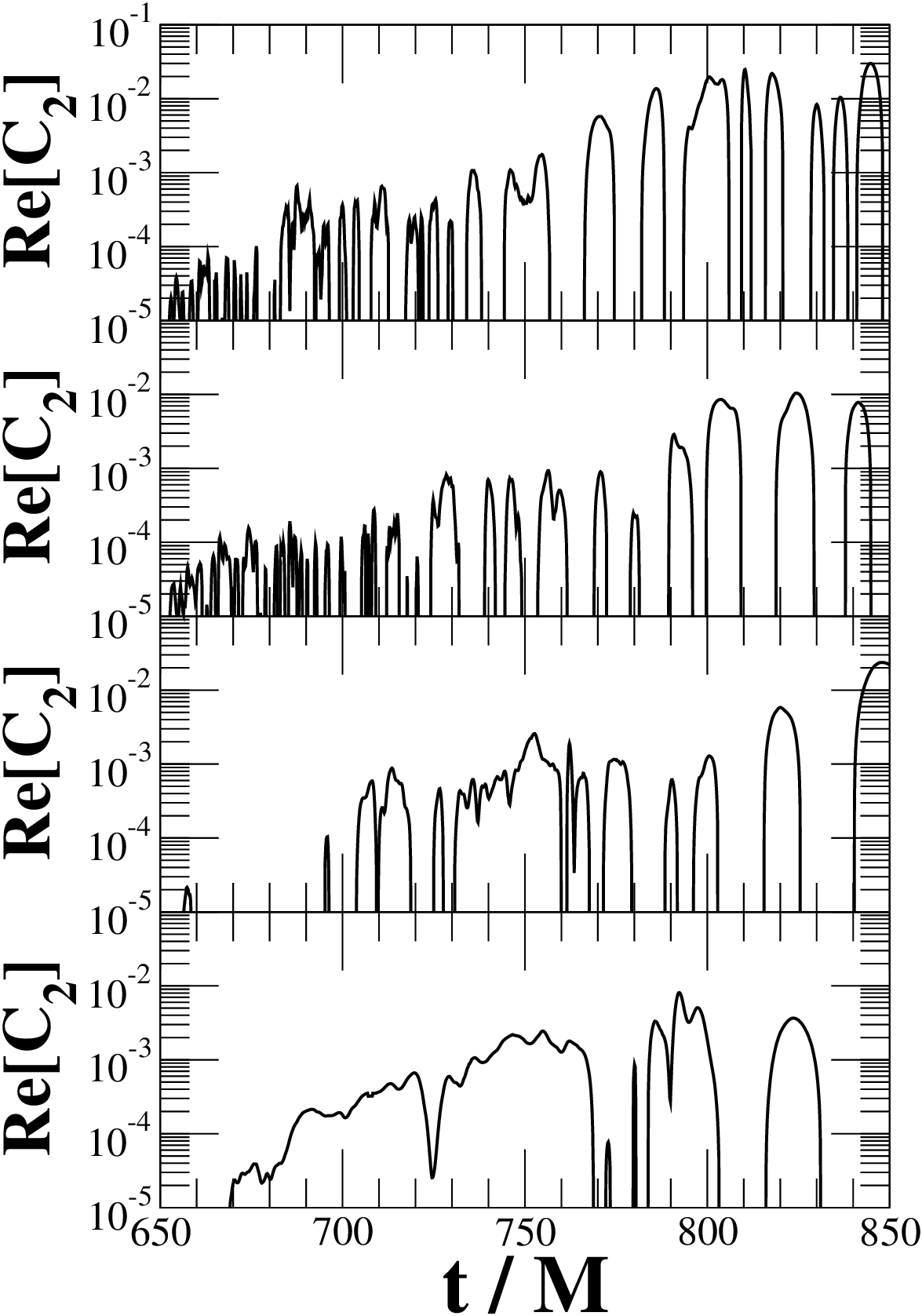}
\caption{
Same as Fig.~\ref{fig:md1rm1} but for the $m=2$ diagnostics.
\label{fig:md1rm2}
}
\end{figure}

\begin{figure}
\centering
\includegraphics[keepaspectratio=true,width=8cm]{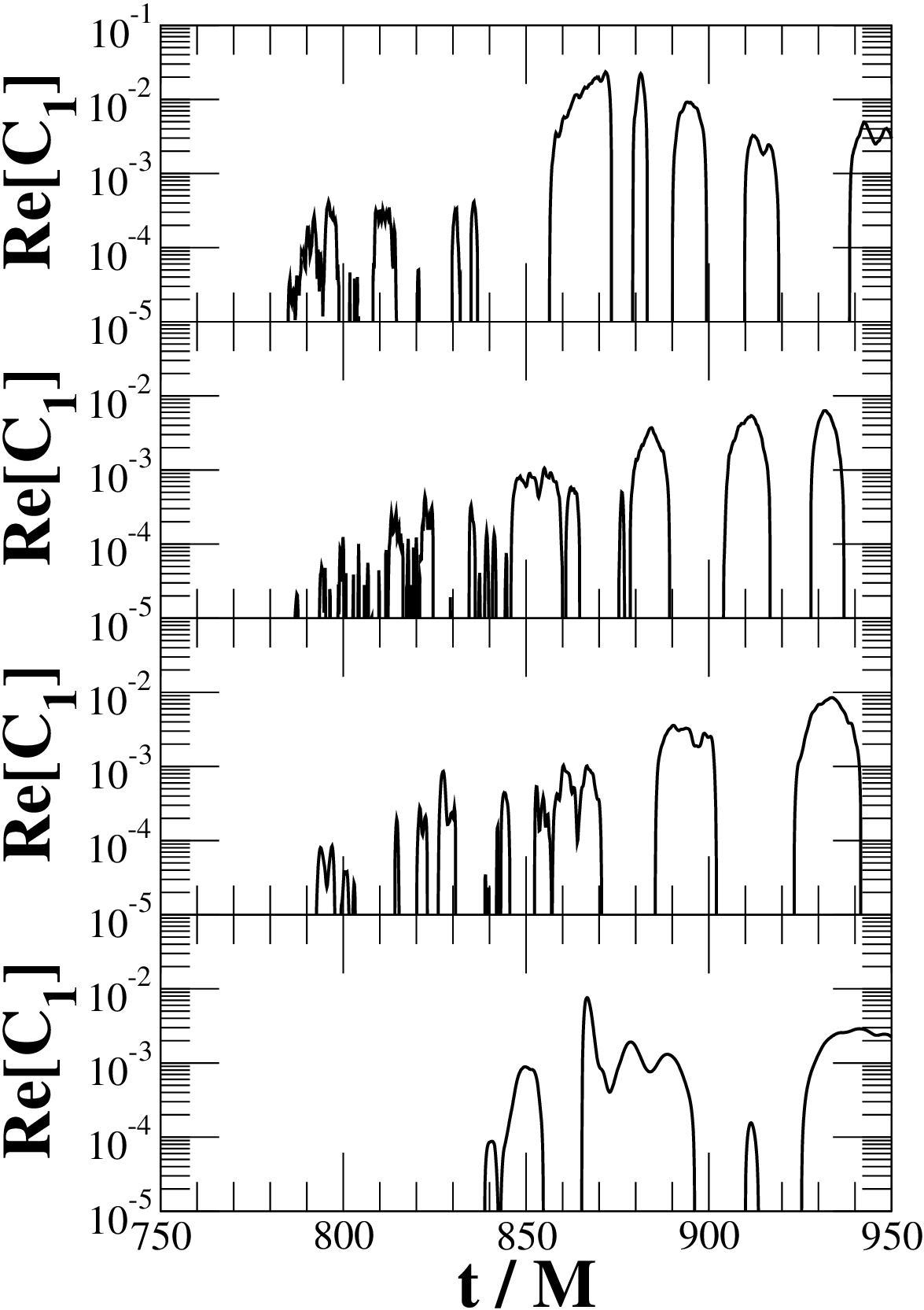}
\caption{
Same as Fig.~\ref{fig:md1rm1} but for model II.  We measure the
diagnostics in the equatorial plane at the radius $r=0.987M$, $1.973
M$, $3.947 M$, $9.867 M$ from the top panel, respectively.
\label{fig:md2rm1}
}
\end{figure}

\begin{figure}
\centering
\includegraphics[keepaspectratio=true,width=8cm]{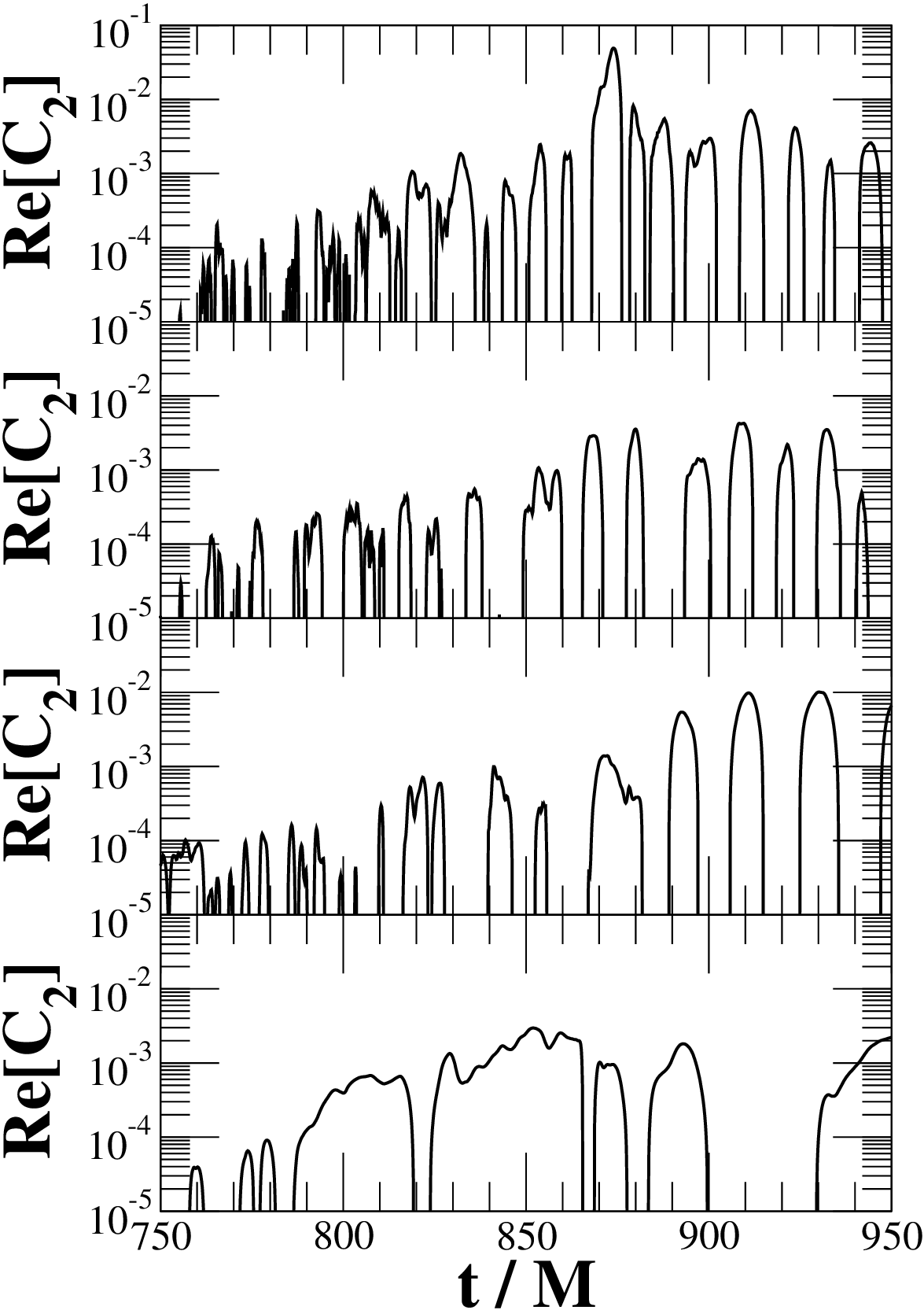}
\caption{
Same as Fig.~\ref{fig:md2rm1} but for the $m=2$ diagnostics.
\label{fig:md2rm2}
}
\end{figure}

\subsection{Horizon configuration}
Here we introduce a useful idea to diagnose the horizon locally in
dynamical spacetime.  It is the dynamical horizon defined as the
outermost trapped tube which is composed of the apparent horizon in
our case \citep{AK03}.  First we have to define the angular momentum
from the horizon configuration.  One way to determine an angular
momentum of the dynamic BH is (see e.g.\ section III.B of
Ref.~\citep{DH06})
\begin{equation}
J_{\rm BH} = - \frac{1}{8 \pi} \int_{S_{\rm R}} K_{\mu \nu} R^{\mu}
\varphi^{\nu} ds,
\end{equation}
where $R^{\mu}$ is the outward directed spacelike normal to the
horizon in the spacelike slice, and $\varphi^{a}$ is a rotational
vector field on the horizon. This quantity is interpreted as the Komar
angular momentum when $\varphi^{a}$ is a rotational Killing vector on
the horizon. The code and method used to compute numerically an
approximate Killing vector, should it exist, is described in
Ref.~\citep{DH03}.

Another way of defining an angular momentum, which is dipole moment of
the scalar curvature $\Im \Psi_{2}$, is 
\begin{equation*}
J_{\rm BH} = - \sqrt{\frac{1}{12\pi}}\frac{A}{4\pi} \int dA \Im
\Psi_{2} Y^{l}_{~0}(u),
\end{equation*}
where $\Psi_{2}$ is Weyl scalar, $Y^{l}_{~m}$ the spherical harmonics,
$A$ the area of the horizon and $u$ the polar angle of the horizon
configuration.  Note that the computations from two different
definitions of the angular momentum coincide with each other in an
axisymmetric apparent horizon.

The dynamical horizon mass of the hole $M_{\rm BH}$ is computed once
we have extracted the angular momentum of the hole by the following
relation 
\begin{equation}
M_{\rm BH} = \frac{1}{2 R_{\rm BH}}\sqrt{R_{\rm BH} + 4 J_{\rm BH}^{2}}.
\end{equation}
Note that $R_{\rm BH}$ is the area radius of the hole.

Axisymmetric isolated horizons are represented by two types of
multipole moment of the scalar curvatures on the apparent horizon as 
\citep{AEPV04}
\begin{eqnarray}
L_{l} &=& - \int \Im \Psi_{2} (u) Y^{l}_{~0}(u) dA,
\\
I_{l} &=& \int \frac{1}{4} {}^{2}{\cal R} (u) Y^{l}_{~0}(u) dA,
\end{eqnarray}
where $\Im \Psi_{2}$ is the imaginary part of the Weyl scalar $\Psi_{2}$
\begin{equation*}
\Psi_{2} = \frac{1}{2} C_{\mu \nu \lambda \rho} 
  (l^{\mu} n^{\nu} l^{\lambda} n^{\rho} - l^{\mu} n^{\nu} m^{\lambda}
\bar{m}^{\rho}),
\end{equation*}
which represents the gravitational monopole at large radius, and 
${}^{2}{\cal R}$ the Ricci scalar.  The quantities $L_{l}$ and $I_{l}$
correspond to the mass and current $l$-pole moment defined in
axisymmetric hole as
\begin{eqnarray}
J_{l} &=& 
  \sqrt{\frac{4\pi}{2l+1}} \frac{R^{l+1}_{\rm BH}}{4 \pi} L_{l},\\
M_{l} &=& 
  \sqrt{\frac{4\pi}{2l+1}} \frac{M_{\rm BH} R^{l}_{\rm BH}}{2 \pi} I_{l}.
\end{eqnarray}
This method has two disadvantage when computing multipole moment
numerically.  One is that the quantities are gauge dependent, and the
other is that they become less accurate in the fixed grid of finite
differencing when computing higher $l$-pole moment.  In order to avoid
the above two issues, we use a different method for computing $l$-pole
moment of the curvatures by introducing averaged quantities on the
trapped surface.  We 
introduce the following ``$n$''-pole moment of ${}^{2}{\cal R}$ and
$\Im \Psi_{2}$ as \citep{Jasiulek09}
\begin{eqnarray}
\mu_{n} ({}^{2}{\cal R}) &=& \langle (\langle {}^{2}{\cal R} \rangle -
   {}^{2}{\cal R})^{n}\rangle,\\
\mu_{n} ({\Im \Psi_{2}}) &=& \langle (\langle {\Im \Psi_{2}} \rangle -
   {\Im \Psi_{2}})^{n} \rangle,
\end{eqnarray}
where the bracket of a physical quantity $\langle Q \rangle$
represents the averaged quantity on the BH horizon 
\begin{eqnarray}
\langle Q \rangle &=& \frac{1}{A} \int Q ~ dA.
\end{eqnarray}
The definition of $n$-pole moment is a general extension of defining
the variance of quantities.  The relations between the quantities
$\mu_{n} ({}^{2}{\cal R})$ and $\mu_{n} ({\Im \Psi_{2}})$, and the mass
and current $l$-pole in axisymmetric spacetime are
\begin{eqnarray}
\mu_{n} ({}^{2}{\cal R}) &=& 
\left\langle 
\left(
  1 - 2 \sum_{l=0}^{\infty} I_{l} Y^{l}_{~0}(u)
\right)^{n}
\right\rangle,
\\
\mu_{n} ({\Im \Psi_{2}}) &=& 
\left\langle 
\left(
  \sum_{l=0}^{\infty} L_{l} Y^{l}_{~0}(u)
\right)^{n}
\right\rangle
.
\end{eqnarray}
Therefore $\mu_{n} ({}^{2}{\cal R})$ corresponds to the summation of
all mass $l$-poles, and $\mu_{n} ({\Im \Psi_{2}})$ all current
$l$-poles.  Since we impose planar symmetry across the equatorial
plane, the current odd $l$-poles vanish entirely.  The disadvantage of
using the quantities $\mu_{n} ({}^{2}{\cal R})$ and $\mu_{n} ({\Im
  \Psi_{2}})$, however, is that it is quite difficult to understand
their physical situation.  Therefore, we introduce the nondimensional
quantities for ${}^{2}{\cal R}$ and $\Im \Psi_{2}$ as
\begin{eqnarray}
{}^{2}{\cal R} & = & \frac{8\pi}{A} {}^{2}{\cal \hat{R}}, \\
{\Im \Psi_{2}} & = & \frac{2\pi}{A} {\Im \hat{\Psi}_{2}},
\end{eqnarray}
from a computational viewpoint.

\section{Numerical Results
\label{sec:NResults}}
Here we focus on the properties of a dynamic BH through gravitational
collapse of a supermassive star.  We choose the equilibrium star
radially unstable for evolution in order to focus on the BH formation
\citep{SH09}.

First we compute the gravitational mass and angular momentum of the
hole with two different definitions for the angular momentum in
Figs.~\ref{fig:mj1} and \ref{fig:mj2}.  One definition for computing
the angular momentum is to use the approximated Killing vector, while
the other is to use dipole moment of the imaginary part of Weyl scalar 
$\Psi_{2}$.  When the dynamical system is axisymmetric, the
computations of the angular momentum by two different definitions
coincides with each other.  We find a clear agreement of the
gravitational mass and angular momentum between two different
definitions in  Fig.~\ref{fig:mj1} for model I and in
Fig.~\ref{fig:mj2} for model II.  The results also tell us that our
gravitational collapse is nearly the same as an axisymmetric dynamics
for both models.

\begin{figure}
\centering
\includegraphics[keepaspectratio=true,width=8cm]{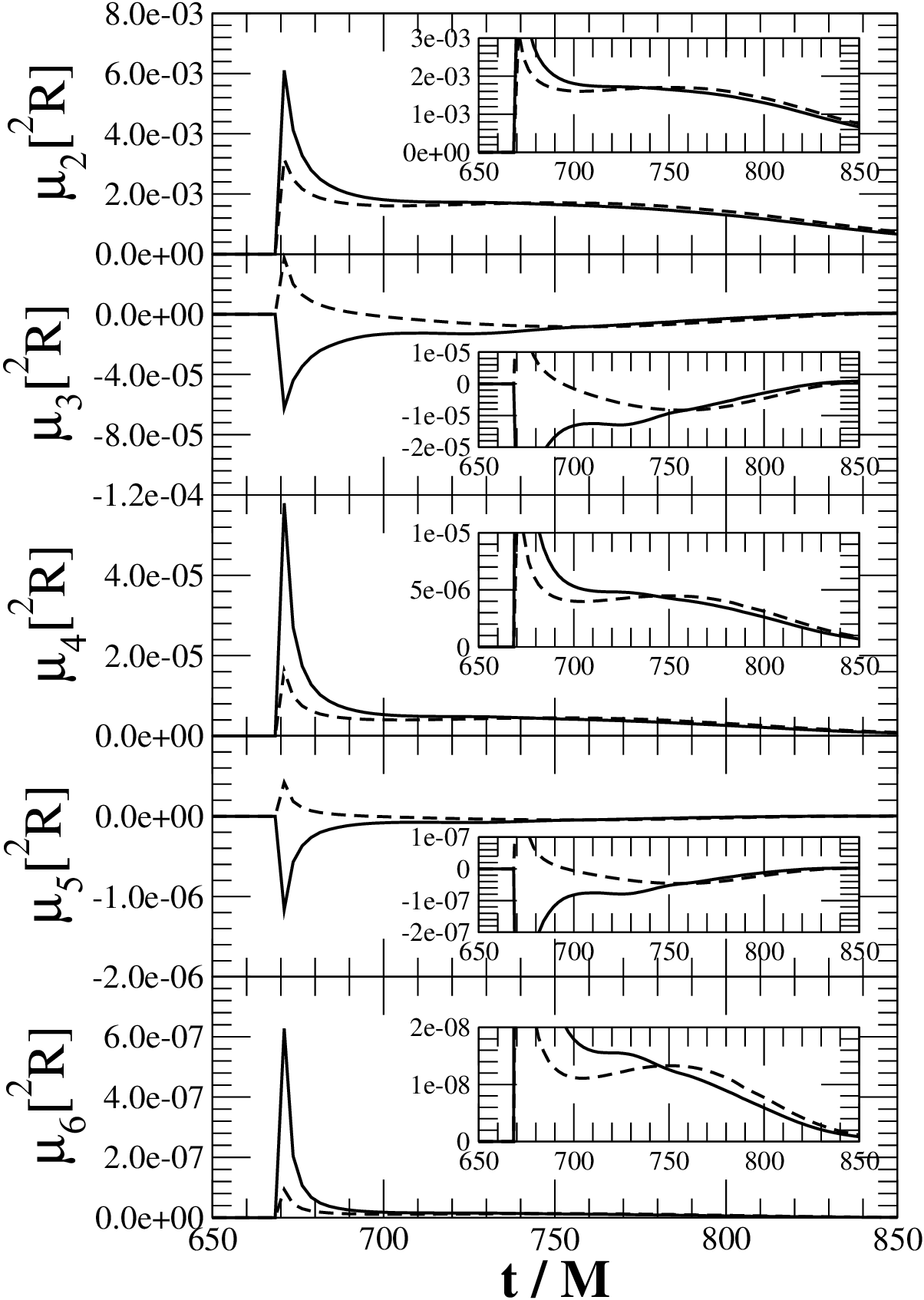}
\caption{
Multipole moment of the Ricci scalar $R$ of the dynamic BH
for model I through evolution.  Solid line represent the one from our 
dynamic BH, while dashed line is the one from a Kerr using the
gravitational mass and angular momentum of the dynamic BH.
We also enlarge the region of quasi-stationary stage at the right part
in each panel.
\label{fig:mm1ra}
}
\end{figure}

\begin{figure}
\centering
\includegraphics[keepaspectratio=true,width=8cm]{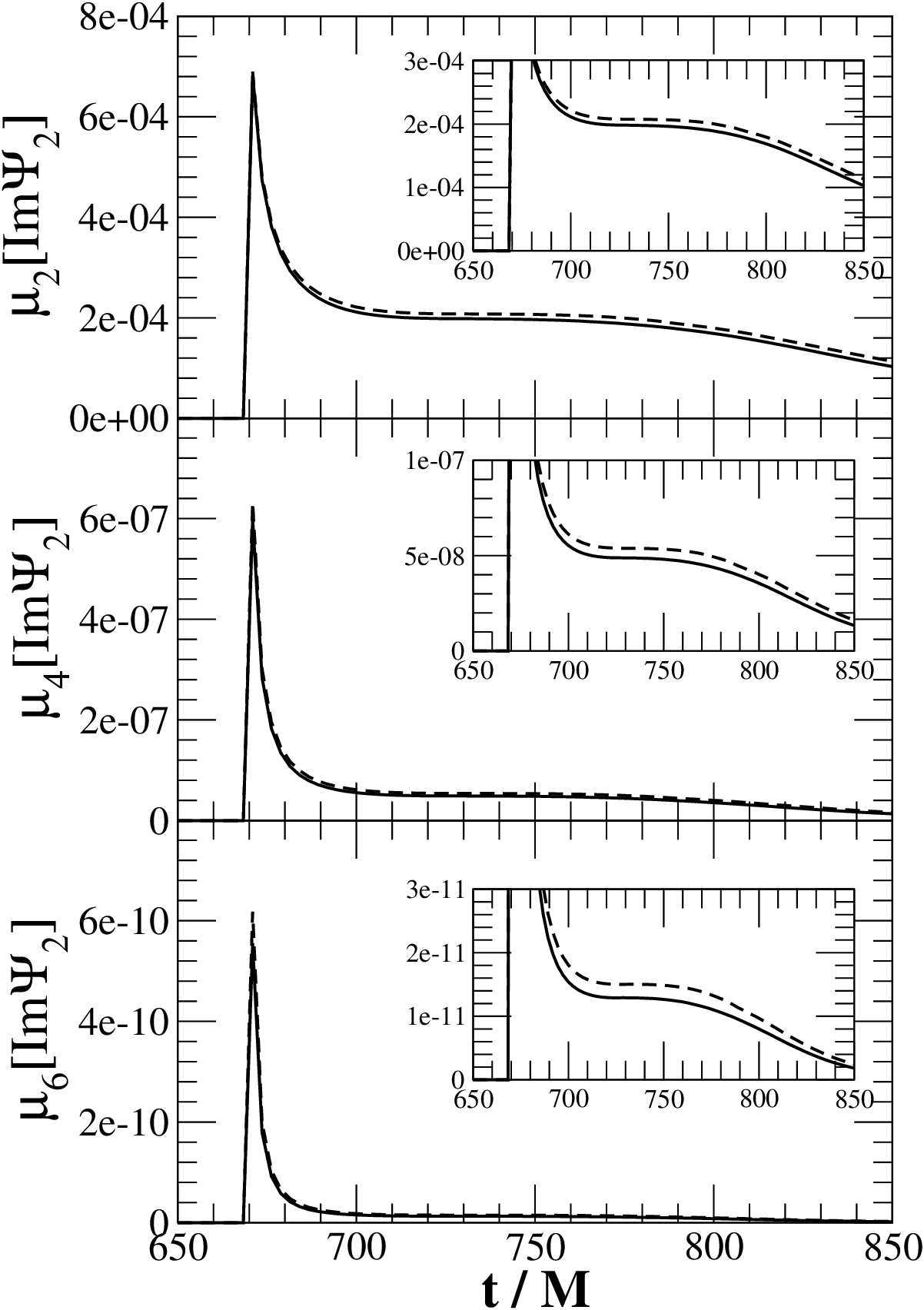}
\caption{
Same as Fig.~\ref{fig:mm1ra} but for the imaginary part of Weyl scalar
$\Psi_{2}$ of the dynamic BH.  Note that the odd $l$-pole of $\Im
\Psi_{2}$ are exactly zero when we impose planner symmetry across the
equator.
\label{fig:mm1ia}
}
\end{figure}

\begin{figure}
\centering
\includegraphics[keepaspectratio=true,width=8cm]{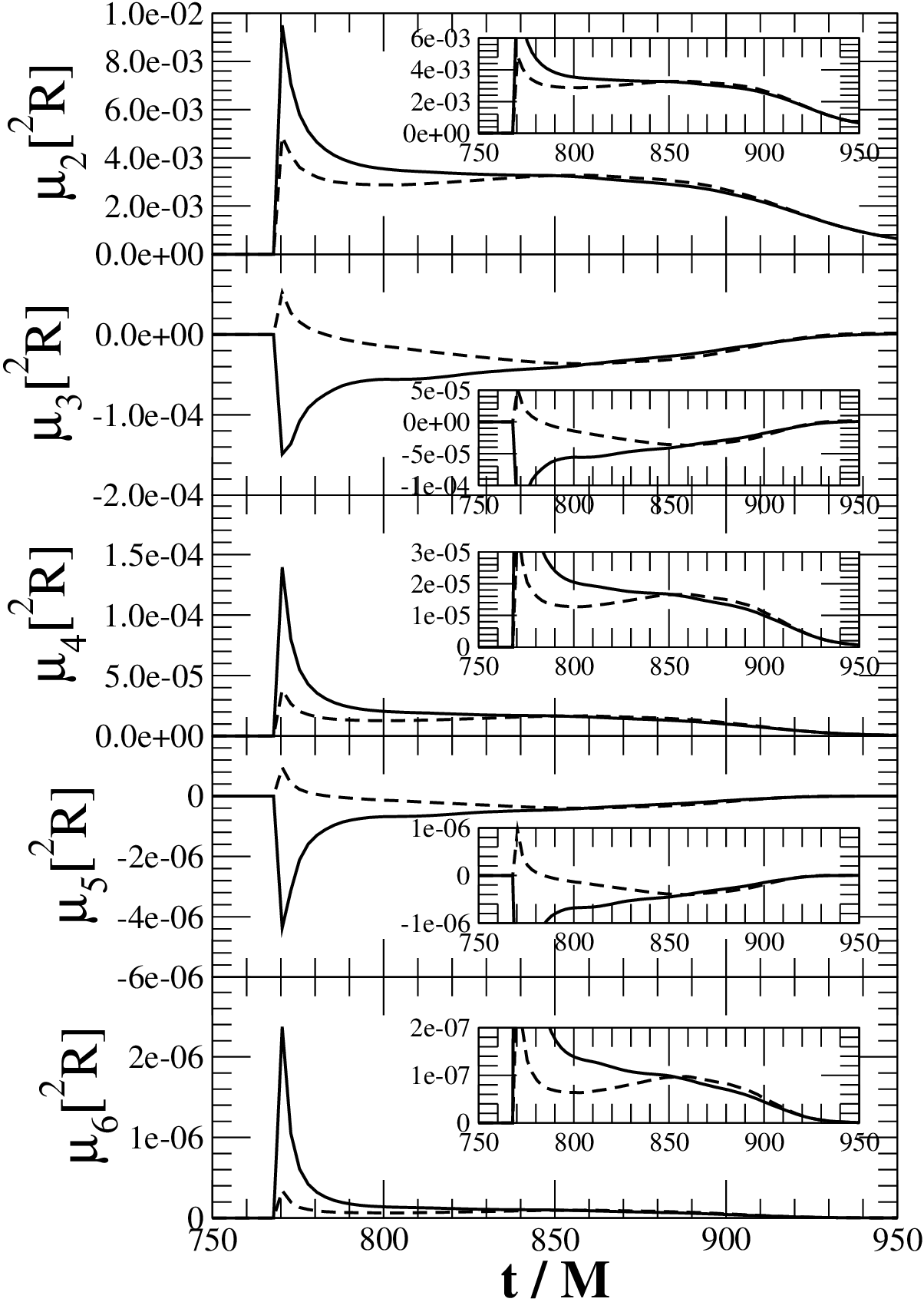}
\caption{
Same as Fig.~\ref{fig:mm1ra} but for model II.
\label{fig:mm2ra}
}
\end{figure}

\begin{figure}
\centering
\includegraphics[keepaspectratio=true,width=8cm]{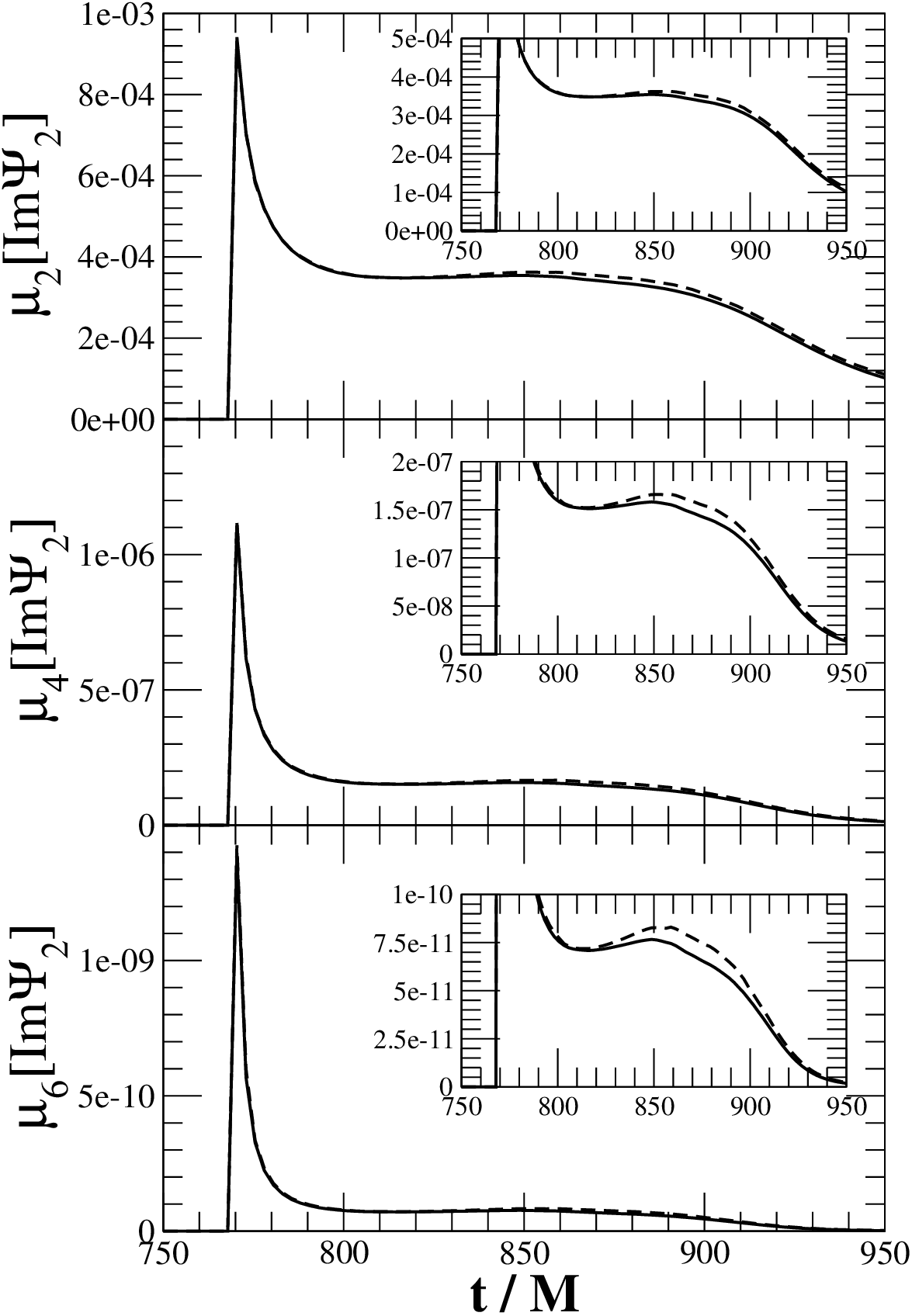}
\caption{
Same as Fig.~\ref{fig:mm1ia} but for model II.
\label{fig:mm2ia}
}
\end{figure}

\begin{figure}
\centering
\includegraphics[keepaspectratio=true,width=8cm]{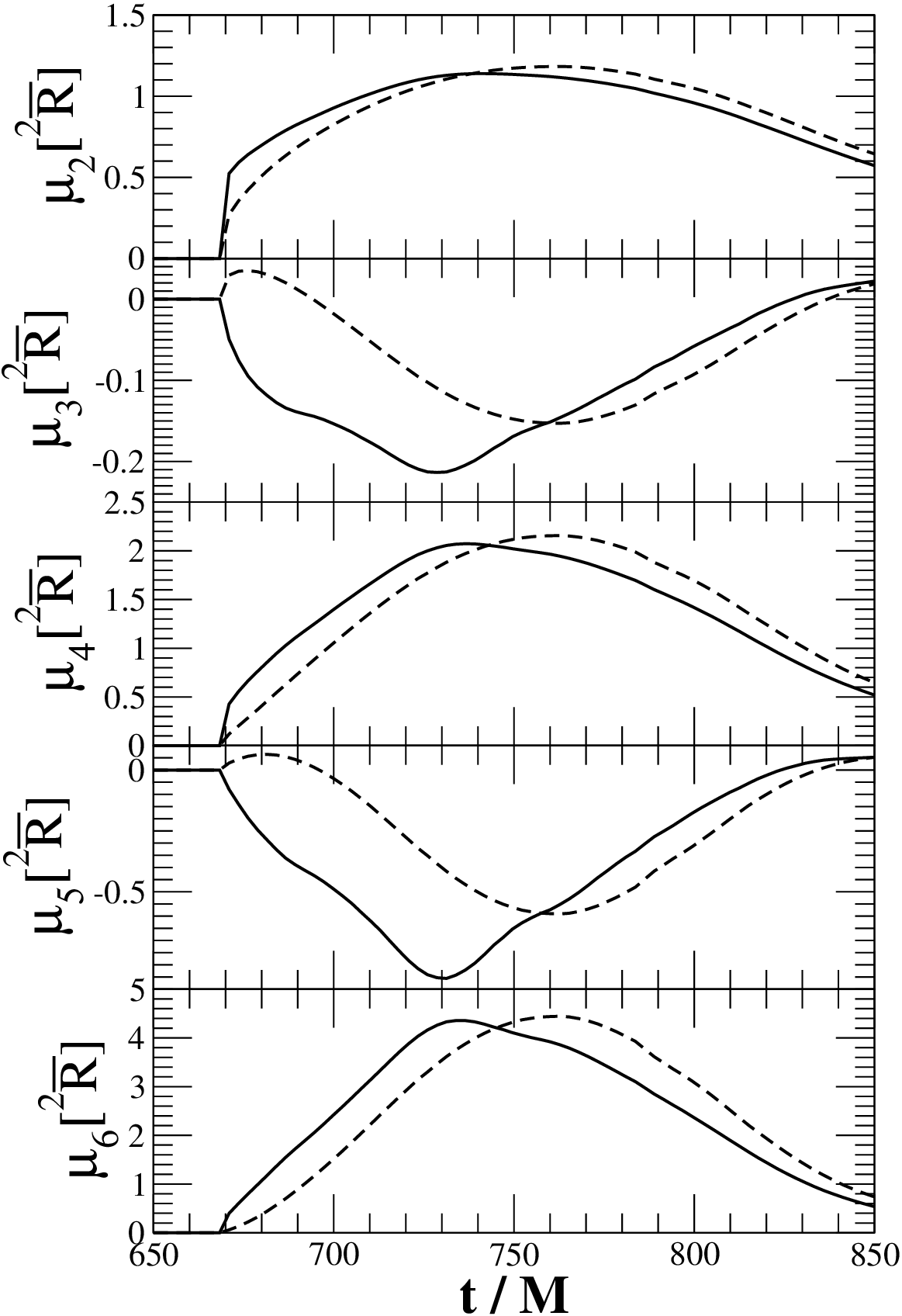}
\caption{
Multipole moment of the Ricci scalar $R$ of the dynamic BH
for model I through evolution.  Solid line represent the one from our 
dynamic BH, while dashed line is the one from a Kerr using the
gravitational mass and angular momentum of the dynamic BH.
\label{fig:mm1r}
}
\end{figure}

\begin{figure}
\centering
\includegraphics[keepaspectratio=true,width=8cm]{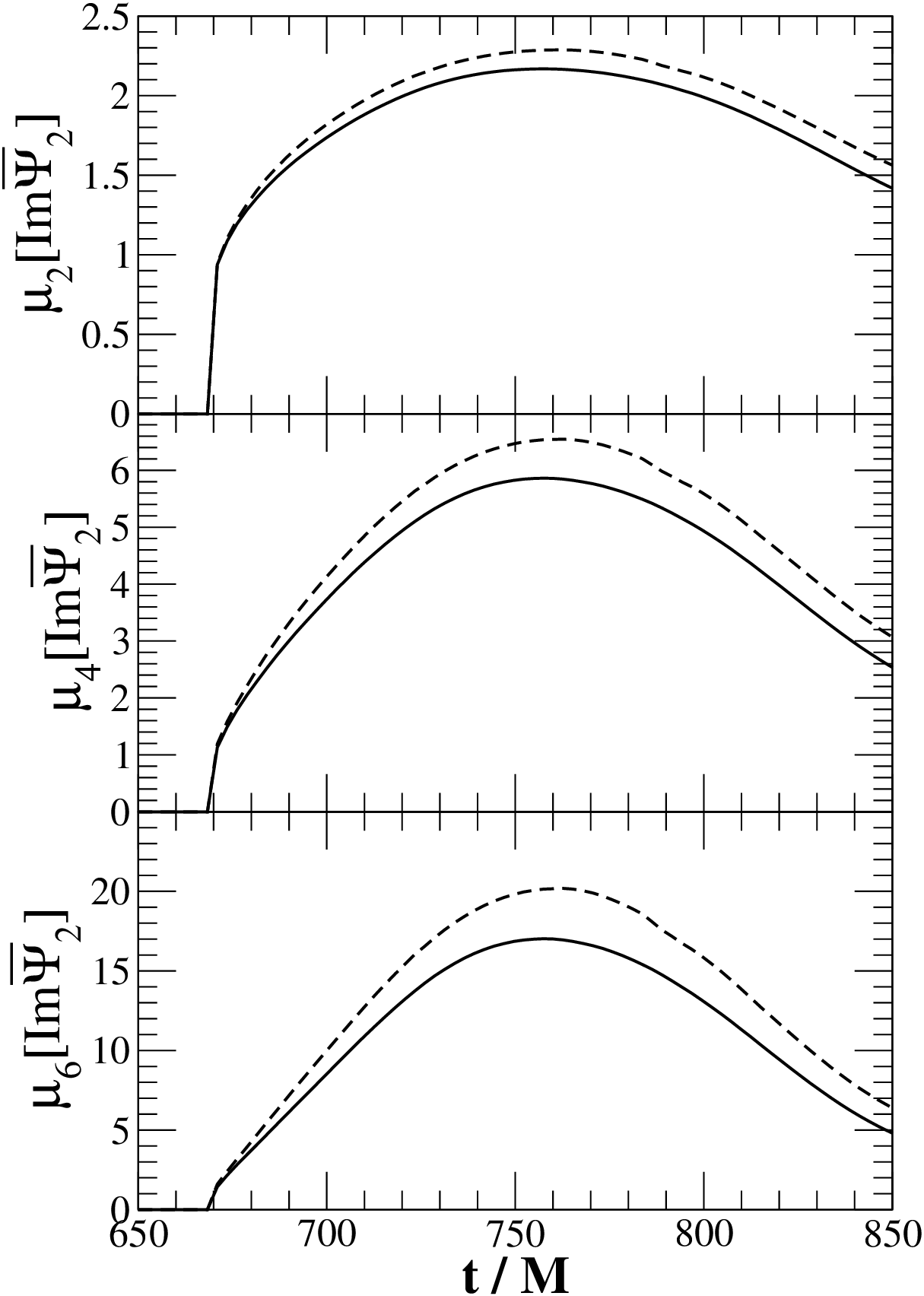}
\caption{
Same as Fig.~\ref{fig:mm1r} but for the imaginary part of Weyl scalar
$\Psi_{2}$ of the dynamic BH.  Note that the odd $l$-pole of $\Im
\Psi_{2}$ are exactly zero when we impose planner symmetry across the
equator.
\label{fig:mm1i}
}
\end{figure}

\begin{figure}
\centering
\includegraphics[keepaspectratio=true,width=8cm]{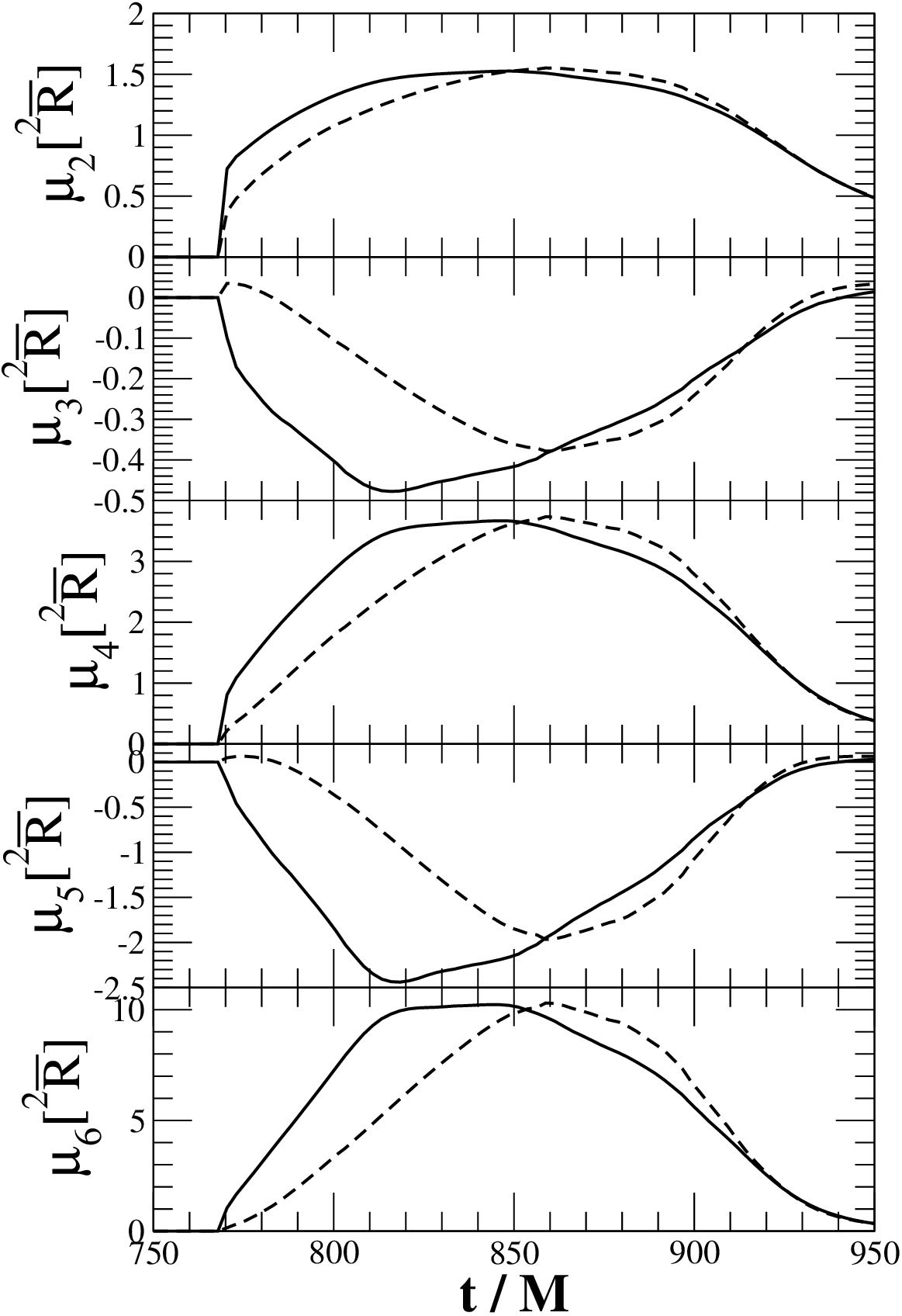}
\caption{
Same as Fig.~\ref{fig:mm1r} but for model II.
\label{fig:mm2r}
}
\end{figure}

\begin{figure}
\centering
\includegraphics[keepaspectratio=true,width=8cm]{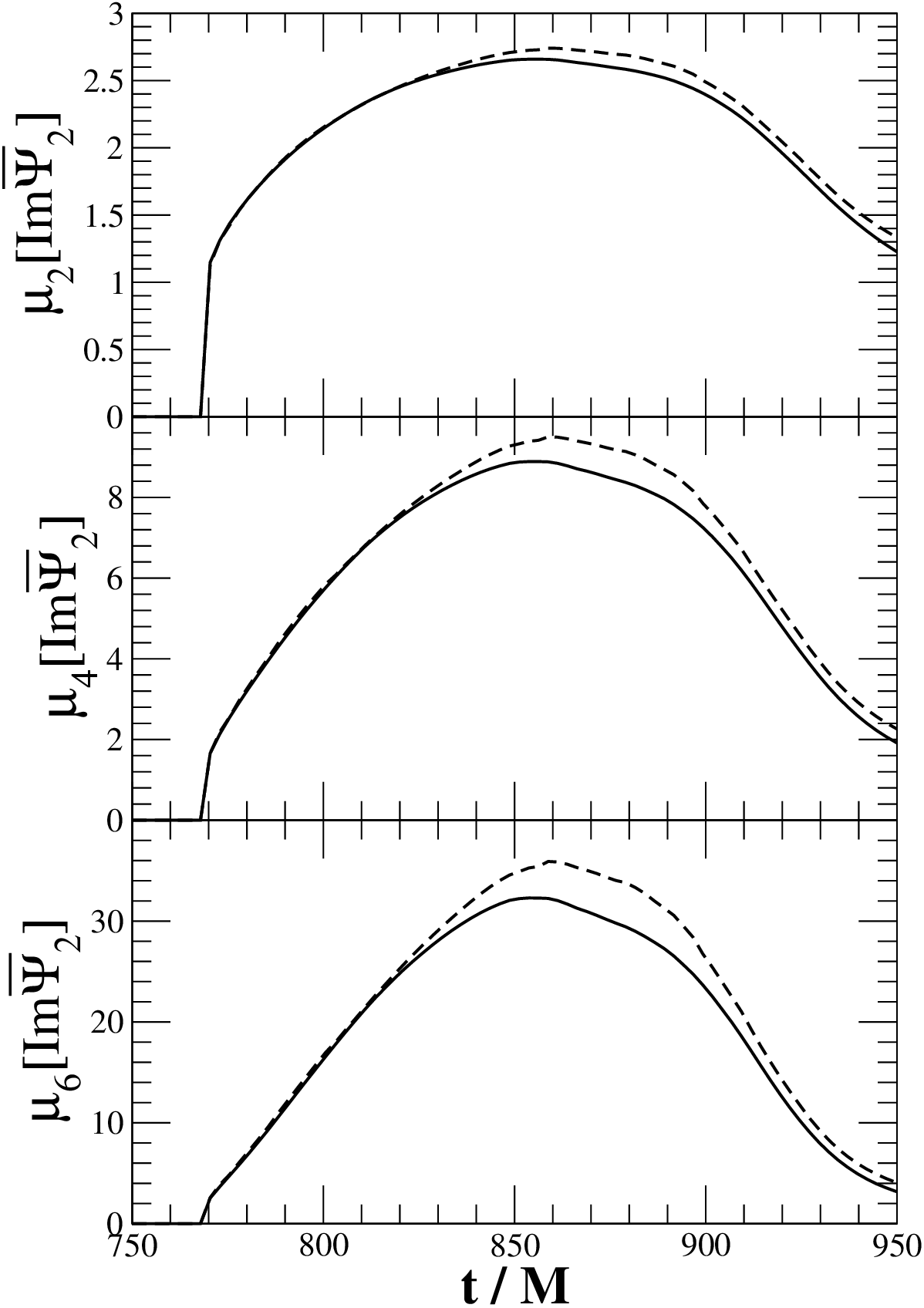}
\caption{
Same as Fig.~\ref{fig:mm1i} but for model II.
\label{fig:mm2i}
}
\end{figure}

Next we monitor the Hamiltonian and Momentum constrains through
gravitational collapse for monitoring the accuracy of our dynamics in
Fig.~\ref{fig:const}.  These checks are necessary because we do not
solve these constraints through time integration of the Einstein's
field equations.  We diagnose the same quantities as before
\citep{SH09}, the Euclidean norm of $HC$ and $MC_{x,y,z}$ of all grid
points outside the apparent horizon, through the normalisation of the
maximum rest mass density outside the horizon in the same
hypersurface.  The maximum violations from the constraints of $MC_{x}$
and $MC_{y}$ are less than $1 \times 10^{-2}$ for model I and $5 
\times 10^{-3}$ for model II.  These facts tell us that our
computational results are very accurate.  They are roughly less than
around $1\%$ relative error at their worst.  However, all the relative
violation errors from the constraints for model I increase at the very
late time of evolution ($t \approx 900 M \sim 1000M$).  We therefore
stop our time integration at the time around $t \approx 1000 M$ to
guarantee roughly $1$\% relative error or less.

We show gravitational waveforms using the Weyl scalar $\Psi_{4}$
through gravitational collapse in Figs.~\ref{fig:gw1} and
\ref{fig:gw2}.  The Weyl scalar $\Psi_{4}$ contains both outgoing
waves and back-scattered waves by the curvature when we measure the
quantity in finite radius from the centre.  In order to focus on the
outgoing waves, we monitor the waveform at three different locations,
and investigate all of them.  Since all three locations are considered
as radiation zone of gravitational waves from the source, the
gravitational field of all three locations is weak, and back-scattered
waves only play a secondary role.  The outgoing waves propagate
towards spatial infinity as time goes on, the features of the outgoing
waves can be seen in all three locations with positive time shift.  We
adjust the time axis of all three waveforms by assuming that
gravitational waves propagate with the speed of light, and plot them
in the same panel in Figs. \ref{fig:gw1comp} and \ref{fig:gw2comp}.
In fact, we use the following relation $t_{\rm adj} \equiv t + x_{\rm
  farthest~observer} - x_{\rm obs}$ to adjust the time.  Although
there is some difference in the magnitude of the amplitude, the 
global features look the same.  Investigating all three waveforms, we
find that the outgoing waves contain three features.  The first is
that there appears a burst wave as the collapse goes on.  The second is
that once the BH forms, there is a damping wave which corresponds to a 
characteristic oscillation of the dynamic BH.  The third is that there
is a continuous wave after the damping one.

In order to identify the cause of continuous waves after the ringdown,
we first investigate the azimuthal modes of the rest mass density.  We
introduce the following diagnostics at certain radii of a ring in the
equatorial plane as
\begin{equation*}
C_{m} = \frac{1}{2\pi D_{\rm ring}} \int_{0}^{2\pi} \rho e^{im\varphi}~d\varphi, 
\end{equation*}
with a normalisation of $D_{\rm ring} (\equiv C_{0})$, a mean density
of the ring at certain radii in the equatorial plane.  We investigate
$m=1$ and $m=2$ diagnostics at 4 different radii for models I and II
in Figs.~\ref{fig:md1rm1} -- \ref{fig:md2rm2}.  Although the
saturation amplitude for different radii is different for each $m$
diagnostic, we find the following features.  The azimuthal diagnostics
begin to amplify efficiently after the apparent horizon has appeared
in the hypersurface.  This feature raises a question as to whether the 
amplification of the azimuthal diagnostics is directly connected with
the configuration of the BH.  The saturation amplitude of each $m$
diagnostics is quite similar at the same radius.  The saturation
amplitude decreases as the radius becomes far from the BH.  This
feature suggests that the matter which is very close to the BH may
play a key role for generating the quasi-periodic waveform after the
ringdown.

Next, we investigate the BH configuration to identify a possible cause
of the continuous waves.  We compute $n$-pole moment of the Ricci
scalar ${\cal R}$ and the imaginary part of the Weyl curvature
$\Psi_{2}$ on the apparent horizon throughout the evolution.  We also
compute $n$-pole moment of the same scalar curvatures using the
configuration of a Kerr BH.  We use the area of the horizon and the
nondimensional Kerr parameter $J_{\rm BH}/ M_{\rm BH}^{2}$ for
computing $n$-pole moment.  Then, we compare each $n$-pole in both
dimensional definition with the BH area in Figs.~\ref{fig:mm1ra} --
\ref{fig:mm2ia}, and nondimensional one in Figs.~\ref{fig:mm1r} --
\ref{fig:mm2i}.  We find the following two features.  The first is
that after $t \approx 100M$ time from the formation of the dynamic
BH \footnote{We define the formation time of the BH as the first
  existent one of the apparent horizon in our hypersurface.}, it is 
described as a Kerr BH is within a relative error of several percents.
If we take the dimension of $n$-poles into account through the area of 
the BH, $n$-pole of the dynamic BH approaches the one of a Kerr in
Figs.~\ref{fig:mm1ra} and \ref{fig:mm1ia} after $t=750M$, and in
Figs.~\ref{fig:mm2ra} and \ref{fig:mm2ia} after $t=850M$.  Therefore
the BH configuration becomes nearly the same as a Kerr after $t\approx
100M$ from the BH formation.  This statement suggests that the cause of
continuous waves may be related to the matter instability, since the
BH configuration is nearly the same as a Kerr.  The other is that the
odd $n$-pole moment has large deviation from that of a Kerr.  This may
be accepted fact as the nonaxisymmetric configuration of the rotating
BH, as it traces the violent phenomenon at the BH formation.  One
caution from this feature is that the BH mass and angular momentum are
settled down after $t \approx 100M$ from the BH formation time (See
Figs.~\ref{fig:mj1} and \ref{fig:mj2}), that the BH is almost regarded
as a Kerr.  For example, the half-life period of the BH oscillation
$\tau (\equiv 1 / \Im \omega_{\rm qnm})$ is $13.6 M$, since the
quasinormal mode frequency of a Kerr BH of $a/M=0.98$ is $M
\omega_{\rm qnm} = 0.422 + 0.0735 i$ for $l=2$, $m=0$
\cite{Leaver85}.  The fact leads to the conclusion that we cannot
extract the ``stationary'' mass and angular momentum of the dynamic BH
by quasinormal ringing in principle.  Those ringing waves represent 
vibration of a transient dynamic BH, not a ``stationary'' one.

\section{Conclusions
\label{sec:Conclusions}}
We investigate the formation of the dynamic BH through gravitational 
collapse by means of three-dimensional hydrodynamic simulations in
general relativity.  We particularly focus on the configuration of a 
dynamic BH and find the following two features.

We investigate two different definitions for the angular momentum of 
a dynamic BH in order to check the validity of the approximated
Killing vector approach.  We compare two results from two different 
definitions for the angular momentum and find that we are able to
extract precisely the BH mass and angular momentum even if we use the
approximated Killing vector.  The fact also indicates that our cases
of gravitational collapse are very close to axisymmetric.

We also demonstrate the method to extract $n$-pole moment of the
dynamic BH precisely without using the approximated Killing vector.
This finding opens a new field of investigating the BH itself by
extracting the properties of $n$-poles of the curvatures on the
horizon.  We compare the configuration of the dynamic BH with that of
the Kerr, using multipole moment of the curvatures on the horizon.  We 
find, as a result, that the quasistationary stage of the newly formed
BH is approximately described by a Kerr BH.  This does not mean,
however, that the whole spacetime is approximately represented by a
Kerr, since we only investigate the trapped surface of the horizon,
just a local structure of the whole spacetime.

\acknowledgments
It is our pleasure to thank Toni Font, Eric Gourgoulhon and Nicolas
Vasset for fruitful discussions.  This work was supported in part by
the JSPS Excellent Young Researcher Overseas Visit Program 2010 and by
the Special Fund for Research program 2009 in Rikkyo University.
Numerical computations were performed on the cluster in the Institute
of Theoretical Physics, Rikkyo University.

\appendix
\section{Multipole moment of the curvatures on the horizon in Kerr
  spacetime}
The 2-surface on the horizon of the Kerr metric in Boyer-Lindquist
coordinate is given as 
\begin{eqnarray}
{}^{2}ds^{2} &=& \Sigma_{+} d\theta^{2} + \frac{(r_{+}^{2} +
  a^{2})^{2}}{\Sigma_{+}} \sin^{2}\theta d\varphi^{2}
\nonumber \\
&=& \eta^{2} [f^{-1}(u) du^{2} + f(u) d\varphi^{2}],
\end{eqnarray}
where 
\begin{eqnarray}
\Sigma_{+} &=& r_{+}^{2} + a^{2} \cos^{2}\theta,\\
r_{+} &=& M + \sqrt{M^{2} - a^{2}}, \\
\eta &=& (r_{+}^{2} + a^{2})^{1/2}, \\
\beta &=& a (r_{+}^{2} + a^{2})^{-1/2}, \\
u &=& \cos\theta, \\
f(u) &=& \frac{1-u^{2}}{1-\beta(1-u^{2})}. 
\end{eqnarray}
The quantities ${}^{2}{\cal R}$ and $\Im \Psi_{2}$ of the Kerr BH on
the horizon are 
\begin{eqnarray}
{}^{2}{\cal R} &=& - \frac{8\pi}{A} \frac{d^{2}}{du^{2}}
\left[ \frac{(1+\hat{c}^{2})(1-u^{2})}{2 (1+\hat{c}^{2} u^{2})} \right], \\
\Im \Psi_{2} &=& \frac{2\pi}{A} \frac{d^{2}}{du^{2}}
\left[ \frac{u (1+\hat{c}^{2})^{2}}{2 \hat{c} (1+\hat{c}^{2} u^{2})}
\right] 
,
\end{eqnarray}
where
\begin{equation}
\hat{c} = \frac{1-\sqrt{1-\hat{a}^{2}}}{1+\sqrt{1-\hat{a}^{2}}},
\end{equation}
and $\hat{a}$ the nondimensional Kerr parameter.  In order to compute
multipole moment of the horizon only from the nondimensional Kerr
parameter and the horizon configuration, we use the nondimensional
quantities of the curvatures ${}^{2}\hat{\cal R}$ and $\Im
\hat{\Psi}_{2}$.

Then, we can compute multipole moment of the Kerr horizon analytically
as 
\begin{widetext}
\begin{eqnarray}
\mu_{2}({}^{2}\hat{\cal R}) &=&
  \frac{-15 - 70 \hat{c}^2 + 128 \hat{c}^4 + 70 \hat{c}^6 + 15
  \hat{c}^8}{80 (1 + \hat{c}^2)} + 
  \frac{3}{16}  (1 + \hat{c}^2)^4 \frac{\arctan \hat{c}}{\hat{c}}
, \\
\mu_{3}({}^{2}\hat{\cal R}) &=&
  \frac{1125 + 5745 \hat{c}^2 + 10965 \hat{c}^4 - 1399 \hat{c}^6 +
    6999 \hat{c}^8 +  6603 \hat{c}^{10} + 2415 \hat{c}^{12} + 315
  \hat{c}^{14}}{2560 (1 + \hat{c}^2)^2} 
\nonumber \\
&& + 
  \frac{9}{512} (1 + \hat{c}^2)^4 (-25 + 14 \hat{c}^2 + 7 \hat{c}^4)
  \frac{\arctan \hat{c}}{\hat{c}}
, \\
\mu_{4}({}^{2}\hat{\cal R}) &=&
  \frac{1}{394240 (1 + \hat{c}^{2})^{3}}
  (-294525 - 1755600 \hat{c}^2 - 4246935 \hat{c}^4 - 5382960
  \hat{c}^6 + 4201406 \hat{c}^8 + 5703728 \hat{c}^{10} + 6818482
  \hat{c}^{12} 
\nonumber \\
&& + 5017584 \hat{c}^{14} + 2127279 \hat{c}^{16} + 480480
  \hat{c}^{18} + 45045 \hat{c}^{20}) 
\nonumber \\
&& +
  \frac{693}{78848} (1 + \hat{c}^{2})^{4} (85 - 60 \hat{c}^2 + 22
  \hat{c}^4 + 52 \hat{c}^6 + 13 \hat{c}^8) 
  \frac{\arctan \hat{c}}{\hat{c}}
, \\
\mu_{5}({}^{2}\hat{\cal R}) &=&
- \frac{1}{7569408 (1 + \hat{c}^2)^4}
  (8437275 + 58760625 \hat{c}^2 + 175151130 \hat{c}^4 + 296175990
  \hat{c}^6 + 325931705 \hat{c}^8 - 152055293 \hat{c}^{10} 
\nonumber \\
&& + 104321660
  \hat{c}^{12} + 411065348 \hat{c}^{14} + 548524189 \hat{c}^{16} +
  426449639 \hat{c}^{18} + 207094602 \hat{c}^{20} + 61939878
  \hat{c}^{22} 
\nonumber \\
&& + 10465455 \hat{c}^{24} + 765765 \hat{c}^{26})
\nonumber \\
&& -
  \frac{3465}{7569408} (1 + \hat{c}^2)^4 (-2435 + 1710 \hat{c}^2 -
  1485 \hat{c}^4 - 572 \hat{c}^6 + 2067 \hat{c}^8 + 1326 \hat{c}^{10}
  + 221 \hat{c}^{12}) 
  \frac{\arctan \hat{c}}{\hat{c}}
, \\
\mu_{6}({}^{2}\hat{\cal R}) &=&
  \frac{1}{8921808896 (1 + \hat{c}^2)^5}
  (-13841202375 - 111349888650 \hat{c}^2 - 396709663350 \hat{c}^4
  - 835111292730 \hat{c}^6 
\nonumber \\
&&  - 1182872163330 \hat{c}^8 - 1230909094610
  \hat{c}^{10} + 906857870914 \hat{c}^{12} + 1101124533086
  \hat{c}^{14} + 1994477265504 \hat{c}^{16} 
\nonumber \\
&& + 3065485548066 \hat{c}^{18} +
  3312605946814 \hat{c}^{20} + 2475795093330 \hat{c}^{22} +
  1279104290178 \hat{c}^{24} + 449618015418 \hat{c}^{26} 
\nonumber \\
&& + 102838870134 \hat{c}^{28} + 13822058250 \hat{c}^{30} + 829323495
  \hat{c}^{32})
\nonumber \\
&& +
  \frac{765765}{8921808896} (1 + \hat{c}^2)^4 (18075 - 11240 \hat{c}^2
  + 15380 \hat{c}^4 - 920 \hat{c}^6 - 10302 \hat{c}^8 + 18216
  \hat{c}^{10} + 23252 \hat{c}^{12} + 8664 \hat{c}^{14} 
\nonumber \\
&& + 1083 \hat{c}^{16})
  \frac{\arctan \hat{c}}{\hat{c}}
, \\
\mu_{2}(\Im \hat{\Psi}_{2}) &=&
  \frac{-15 + 170 \hat{c}^2 + 112 \hat{c}^4 + 70 \hat{c}^6 + 15
  \hat{c}^8}{80 (1+\hat{c}^{2})} + \frac{3}{16} (1+\hat{c}^{2})^{4} 
  \frac{\arctan \hat{c}}{\hat{c}}
, \\
\mu_{4}(\Im \hat{\Psi}_{2}) &=&
  \frac{1}{49280 (1 + \hat{c}^2)^3}
  (-3465 - 36960 \hat{c}^2 + 619773 \hat{c}^4 + 663168 \hat{c}^6
    + 1273910 \hat{c}^8 + 1306240 \hat{c}^{10} + 985930 \hat{c}^{12} +
    515328 \hat{c}^{14} 
\nonumber \\
&& + 178563 \hat{c}^{16} + 36960 \hat{c}^{18} +
    3465 \hat{c}^{20})
\nonumber \\
&& +
  \frac{693}{9856} (1 + \hat{c}^2)^{8} \frac{\arctan \hat{c}}{\hat{c}}
, \\
\mu_{6}(\Im \hat{\Psi}_{2}) &=&
  \frac{1}{8921808896 (1 + \hat{c}^2)^5}
  (-271846575 - 4530776250 \hat{c}^2 - 35485039590 \hat{c}^4 + 
    755725266582 \hat{c}^6 + 749954442094 \hat{c}^8 
\nonumber \\
&& + 2536970204990
    \hat{c}^{10} + 3859089592210 \hat{c}^{12} + 5130563563118
    \hat{c}^{14} + 5335972052992 \hat{c}^{16} + 4426939814290
    \hat{c}^{18} 
\nonumber \\
&& + 2910344048750 \hat{c}^{20} + 1498699974850
    \hat{c}^{22} + 592140524690 \hat{c}^{24} + 173417402730
    \hat{c}^{26} + 35485039590 \hat{c}^{28} 
\nonumber \\
&& + 4530776250 \hat{c}^{30}
    + 271846575 \hat{c}^{32})
\nonumber \\
&& +
   \frac{271846575}{8921808896} (1 + \hat{c}^2)^{12} 
   \frac{\arctan\hat{c}}{\hat{c}}
.
\end{eqnarray}
\end{widetext}


\end{document}